%% file: main.tex
\documentclass[twoside,leqno,twocolumn]{article}

\usepackage[letterpaper]{geometry}

\usepackage{siamproceedings}
\usepackage[T1]{fontenc}

\usepackage{hyperref}
\usepackage{graphicx}
\usepackage{textcomp}
\usepackage{xcolor}

\newif\iffull
\fulltrue

\usepackage{style}
\input{macro}
\usepackage{thmtools}
\usepackage{thm-restate}

\begin{document}

\title{PECANN: Parallel Efficient Clustering with Graph-Based
Approximate Nearest Neighbor Search
}

\author{
Shangdi Yu\thanks{MIT (\email{shangdiy@mit.edu}, \email{jengels@mit.edu}, \email{yh\_huang@mit.edu}, \email{jshun@mit.edu})}
\and Joshua Engels\footnotemark[1]    
\and Yihao Huang\footnotemark[1] 
\and Julian Shun\footnotemark[1]    
}

\date{}

\maketitle

\fancyfoot[R]{\scriptsize{Copyright \textcopyright\ 2025 by SIAM\\
Unauthorized reproduction of this article is prohibited}}

\input{abstract}

\input{intro}

\input{prelim}

\input{framework}

\input{optimization}

\input{usage}

\input{analysis}

\input{experiment}

\input{related_work}
\input{conclusion}

\section*{Acknowledgments.}
This research is supported by MIT PRIMES, Siebel Scholars program,
DOE Early Career Award \#DE-SC0018947,
NSF Awards \#CCF-1845763, \#CCF-2316235, and \#CCF-2403237,
Google Faculty Research Award, Google Research Scholar Award, cloud computing credits from Google-MIT, and
FinTech@CSAIL Initiative. We thank Magdalen Manohar for helping with the ParlayANN code base, and Amartya Shankha Biswas, Ronitt Rubinfeld, and Harsha Simhadri for helpful discussions.

\bibliographystyle{siamplain}
\bibliography{references}

\iffull
\input{appendix-functions}
\input{appendix-analysis}

\input{appendix-exp}

\fi

\end{document}

%% file: macro.tex
\usepackage[utf8]{inputenc}

\usepackage{pgfplots}

\usepackage{amsmath,amssymb, amsfonts}
\usepackage{cleveref}
\usepackage{xy}
\usepackage{tikz}
\usepackage{mathrsfs}
\usepackage{setspace,dsfont,graphicx,makecell,appendix,multirow,float}
\pgfplotsset{compat=1.17}

\usepackage{bbm}
\usepackage{graphicx}

\usepackage{tabularx}
\usepackage{dblfloatfix}
\usepackage{xspace}
\usepackage{algorithm}
\usepackage{subcaption,graphicx,xurl,multicol,mathtools}
\usepackage{paralist}
\usepackage{color}
\usepackage{arydshln}
\usepackage[most]{tcolorbox}
\usepackage{stmaryrd}
\usepackage{booktabs}
\usepackage{siunitx}
\usepackage{bm}
\graphicspath{ {./figures/} }

\algblock{ParFor}{EndParFor}
\algrenewtext{ParFor}[1]{\algorithmicparfor\ #1\ \algorithmicpardo}
\algrenewtext{EndParFor}{\algorithmicendparfor}
\algtext*{EndParFor}

\algdef{SE}[DOWHILE]{Do}{doWhile}{\algorithmicdo}[1]{\algorithmicwhile\ #1}%

\usepackage{comment}

\newcommand{\knn}{$k$\text{-nearest} \text{neighbor}\xspace}
\newcommand{\framework}{PECANN\xspace}

\newcommand{\savespace}[1]{}

\usepackage{enumitem}
\setlist{nosep}

%% file: abstract.tex
\begin{abstract}

In this paper, we study variants of density peaks clustering, a popular type of density-based clustering algorithm for points that has been shown to work well in practice. Our goal is to cluster \emph{large high-dimensional} datasets, which are prevalent in practice. Prior solutions are either sequential and cannot scale to large data, or are specialized for low-dimensional data.
This paper unifies the different variants of density peaks clustering into a single framework, \framework (\textbf{P}arallel \textbf{E}fficient \textbf{C}lustering with \textbf{A}pproximate \textbf{N}earest \textbf{N}eighbors), by abstracting out several key steps common to this class of algorithms. One such key step is to find nearest neighbors that satisfy a predicate function, and
one of the main contributions of this paper is an efficient way to do this predicate search using graph-based approximate nearest neighbor search (ANNS).
To provide ample parallelism, we propose a doubling search technique that enables points to find an approximate nearest neighbor satisfying the predicate in a small number of rounds. 
Our technique can be applied to many existing graph-based ANNS algorithms, which can all be plugged into \framework.

We implement five clustering algorithms with \framework and evaluate them on synthetic and real-world datasets with up to $1.28$ million points and up to $1024$ dimensions on a 30-core machine with two-way hyper-threading. 
Compared to the state-of-the-art \algname{fastdp} algorithm for high-dimensional density peaks clustering, which is sequential, our best algorithm is 45x--734x faster while achieving competitive ARI scores. Compared to the state-of-the-art parallel DPC-based algorithm, which is optimized for low dimensions, \framework is two orders of magnitude faster.
As far as we know, we are the first to evaluate DPC variants on large high-dimensional real-world image and text embedding datasets.
\end{abstract}

%% file: intro.tex
\section{Introduction}\label{sec:intro}

Clustering is the task of grouping similar objects into clusters and is a fundamental task in data analysis and unsupervised machine learning~\cite{Jain1999,Aggarwal2013,berkhin2006survey}. 
For example, clustering algorithms can be used to identify different types of tissues in medical imaging~\cite{Yang02}, analyze social networks~\cite{MishraSST07}, and identify weather regimes in climatology~\cite{Coe21}. They are also widely used as a data processing subroutine in other machine learning tasks~\cite{Coleman79,Wu22,Lin19,Marco13}. One popular type of clustering is density-based clustering, where clusters are defined as dense regions of points in space. Recently, density-based clustering algorithms have received a lot of attention~\cite{Ester96,Agrawal98,Ankerst99,Januzaj04,Rodriguez14,Wang97,Hinneburg98,Hanmanthu18,Sheikholeslami00} because they can discover clusters of arbitrary shapes and detect outliers (unlike popular algorithms such as $k$-means, which can only detect spherical clusters).

Density peaks clustering (DPC)~\cite{Rodriguez14} is a popular density-based clustering technique for spatial data (i.e., point sets) that has proven very effective at clustering challenging datasets with non-spherical clusters. 
Due to DPC's success, many DPC variants have been proposed in the literature (e.g., ~\cite{sddp, chen2020fast, sieranoja2019fast,yaohui2017adaptive, sun2021density, xie2016robust,yin2022improved,du2018density,su2018bpec, hou2019enhancing, geng2018recome}). However, existing DPC variants are sequential and/or tailored to low-dimensional data, and so cannot scale to the large, high-dimensional datasets that are common in practice. 

This paper addresses this gap by proposing 
a novel framework called \framework: \textbf{P}arallel \textbf{E}fficient \textbf{C}lustering with \textbf{A}pproximate \textbf{N}earest \textbf{N}eighbors. \framework contains implementations for a variety of different DPC density techniques that both scale to large datasets (via efficient parallel implementations) and run on high dimensional data (via approximate nearest neighbor search).
Designing a unifying framework for DPC variants is non-trivial, as DPC variants can differ significantly. Developing a modular and extensible framework that can seamlessly incorporate various DPC variants and allow for easy comparison and experimentation requires careful abstraction and encapsulation of the key algorithmic components. 
Furthermore, extending DPC to high dimensions is challenging as there are no efficient parallel solutions for constrained nearest neighbor search in high dimensions, which is needed for DPC.
Before going into more details on our contributions, we review the main steps of DPC variants and discuss existing bottlenecks. 

The three key steps of DPC variants are as follows:
\begin{mdframed}\small
\begin{enumerate}[topsep=1pt,itemsep=0pt,parsep=0pt,leftmargin=8pt,label=(\textbf{\arabic*})]
    \item Compute the density of each point $x$. \label{item:density}
    \item Construct a tree by connecting each point $x$ to its closest neighbor with higher density than $x$. \label{item:dep-point}
    \item Remove edges in the tree according to a pruning heuristic. Each resulting connected component is a separate cluster.\label{item:single-linkage} 
\end{enumerate}
\end{mdframed}

Step~\ref{item:density} is computed differently based on the variant, but all variants use a function that depends on either the $k$-nearest neighbors of $x$ or the points within a given distance from $x$. Efficient implementations of this step rely on nearest neighbor queries or range queries. In low dimensions, these queries can be answered efficiently using spatial trees, such as $k$d-trees. However, $k$d-trees are inefficient in high dimensions due to the curse of dimensionality~\cite{weber1998quantitative}. 
Step~\ref{item:dep-point} again requires finding nearest neighbors, but with the constraint that only neighbors with higher density are considered. 
Step~\ref{item:single-linkage} can easily be computed using any connected components algorithm. 
Steps~\ref{item:density} and~\ref{item:dep-point} form the bottleneck of the computation, and take quadratic work in the worst case, while Step~\ref{item:single-linkage} can be done in (near) linear work. 
Note that different clusterings can be generated by
reusing the tree from Step~\ref{item:dep-point} and
simply re-running Step~\ref{item:single-linkage} using different pruning heuristics. 
The tree from Step~\ref{item:dep-point} can be viewed as a cluster hierarchy (or dendrogram) that contains clusterings at different resolutions.

Existing papers on DPC variants mainly focus on their own proposed variant, and as far as we know, there is no unified framework for implementing and comparing DPC variants and evaluating them on the same datasets. Furthermore, most DPC papers focus on clustering low-dimensional data, but many datasets in practice are high dimensional ($d > 100$). 
The \framework framework unifies a broad class of DPC variants by abstracting out these three steps and providing efficient parallel implementations for different variants of each step.  
For Step~\ref{item:density}, we leverage graph-based approximate ANNS algorithms, which are fast and accurate in high dimensions~\cite{ANNScaling, wang2021comprehensive}. 
For Step~\ref{item:dep-point}, we adapt graph-based ANNS algorithms to find higher density neighbors by iteratively doubling the number of nearest neighbors returned until finding one that has higher density. Our doubling search guarantees that the algorithm finishes in a logarithmic number of rounds, making it highly parallel. 
For Steps~\ref{item:density} and~\ref{item:dep-point}, \framework supports the following graph-based ANNS algorithms: \algname{Vamana}~\cite{jayaram2019diskann}, \algname{pyNNDescent}~\cite{pynndescent}, and \algname{HCNNG}~\cite{munoz2019hcnng}.
For Step~\ref{item:single-linkage}, we use a concurrent union-find algorithm~\cite{Jayanti21} to achieve high parallelism.
Prior work~\cite{sieranoja2019fast} has explored using graph-based ANNS for high-dimensional clustering, but their algorithm is not parallel and they only consider one DPC variant and one underlying ANNS algorithm. 
In addition, we provide theoretical work and span bounds
of \framework that depend on the complexity of the underlying ANNS algorithm.
\framework is implemented in C++, using the ParlayLib~\cite{Blelloch20} and ParlayANN~\cite{ANNScaling} libraries, and also has Python bindings.

We use \framework to implement five DPC variants and evaluate them on a variety of synthetic and real-world data sets with up to
$1.28$ million points and up to $1024$ dimensions. We find that using a density function that is the inverse of the distance to the $k^\text{th}$ nearest neighbor, combined with the \algname{Vamana} algorithm for ANNS, gives the best overall performance.
On a 30-core machine with two-way hyper-threading, this best algorithm in \framework achieves 
37.7--854.3x speedup over a parallel brute force approach, and 45--734x speedup over \algname{fastdp}~\cite{sieranoja2019fast}, the state-of-the-art DPC-based algorithm for high dimensions, while achieving similar accuracy in terms of ARI score. \algname{fastdp} is sequential, but even if we assume that it achieves a perfect speedup of 60x, \framework still achieves a speedup of 0.76--12.24x.
Compared to the state-of-the-art parallel density peaks clustering algorithm by Huang et al.~\cite{dpc}, which is optimized for low dimensions, our best algorithm achieves a 320x speedup while achieving a higher ARI score
on the MNIST dataset (their algorithm failed on larger datasets). 

Our contributions are summarized below.

\begin{enumerate}[topsep=1pt,itemsep=0pt,parsep=0pt,leftmargin=10pt,label=\arabic*.]
    \item We introduce the \framework framework 
    that unifies existing \knn-based DPC variants and supports parallel implementations of them that scale to large high-dimensional datasets. We provide fast parallel implementations for five DPC variants. 
    \item We extend graph-based ANNS algorithms with a parallel doubling-search method for finding higher density neighbors. 
    \item We perform comprehensive experiments on a 30-core machine with two-way hyper-threading showing that \framework 
    outperforms the state-of-the-art DPC-based algorithm for high dimensions by 45--734x. As far as we know, we are the first to compare different variants of DPC on large high-dimensional real-world image and text embedding datasets. 
\end{enumerate}

Our code and the full version of the paper are available at \url{https://github.com/yushangdi/PECANN-DPC}.

%% file: prelim.tex
\section{Preliminaries}\label{sec:prelim}
\subsection{Definitions and Notation}
A summary of the notation is provided in \Cref{tab:notation}. Let $P = \{x_1, \ldots, x_n\}$ represent a set of $n$ points in $d$-dimensional coordinate space to be clustered. We use 
$x_i$ to represent the $i^\text{th}$ point in $P$. 
Let $G$ be a search index that supports searching for the exact or approximate nearest neighbors of a query point.
Let $D(x_i, x_j)$ denote the distance (dissimilarity) between points $x_i$ and $x_j$, where a larger distance value means the points are less similar. $D$ can be any distance measure the search index $G$ supports. 

\begin{table}[t]
  \centering
  \vspace{-3pt}
  \footnotesize
  \begin{tabular}{cc}
   \toprule
   Notation & Meaning \\
   \midrule
    $P$ & input set of points \\
    $n$, $d$ & size and dimensionality of $P$ \\
    $x_i$ &  $i^\text{th}$ point in $P$ \\
    $G$ & a similarity search index \\
    $D(x_i, x_j)$ & distance (dissimilarity) between $x_i$ and $x_j$\\
    $\rho_i$, $\lambda_i$ & density and dependent point of $x_i$ \\
    $\delta_i$ & dependent distance of $x_i$ (i.e., $D(x_i,\lambda_i)$) \\
    $k$ & the number of neighbors used for computing densities\\
    $\mathcal{N}_i$ & (approximate) $k$-nearest neighbors of $x_i$ \\
    $\mathcal{W}_c, \mathcal{S}_c$ & the work and span of constructing $G$\\
    $\mathcal{W}_{nn}, \mathcal{S}_{nn}$ & the work and span of finding nearest neighbors using $G$\\
    \bottomrule
  \end{tabular}
  \caption{Notation}
  \label{tab:notation}
  \vspace{-5pt}
\end{table}

Let the \defn{neighbors} ($\mathcal{N}_i$) of a point $x_i$ be either its exact or approximate $k$-nearest neighbors.
Let $\rho_i$ be the \defn{density} of point $x_i$, representing how dense the local region around $x_i$ is. A larger $\rho_i$ value indicates a denser local region. 
For example, in the original DPC algorithm~\cite{Rodriguez14}, the density of a point $x$ is the number of points within a given radius of $x$, and  
in the SD-DP (sparse dual of density peaks) algorithm~\cite{sddp}, the density of a point is the inverse of its distance to its $k^\text{th}$ nearest neighbor. In this paper, we consider the densities that can be computed from the \knn{s} of $x$.

\begin{definition}\label{def:dep-pt}
   Let $P_i = \{x_j \mid x_j\in P \wedge \rho_j >\rho_i\}$.
    For $x_i$, its exact \textbf{dependent point} is a point $\lambda_i \in P_i$ such that,
    $
    D(x_i,\lambda_i) \le D(x_i, x_j)\ \forall \ x_j \in P_i 
    $ (i.e., it is the closest point with higher density than $x_i$).
    The \textbf{dependent distance} ($\delta_i$) of $x_i$ is $D(x_i,\lambda_i)$, i.e., the distance to its dependent point (or $\infty$ if it does not have one). 
\end{definition}

\Cref{def:dep-pt} defines the dependent point to be the \emph{closest} point with higher density, which is expensive to compute in high dimensions. For high-dimensional data, we relax the constraint to allow reporting an \emph{approximate} nearest neighbor with higher density (i.e., considering just the points with higher density, choose approximately the closest one). 
Roughly speaking, an \defn{approximate nearest neighbor} of a point $x$ is one whose distance from $x$ is not too far from the distance of the true nearest neighbor from $x$.
In our experiments, we use the Euclidean distance function, one of the most commonly used distance functions for clustering.

Points that are outliers and do not belong to any cluster are classified as \defn{noise points}. A noise point is in its own singleton cluster.
For example, some algorithms require a density cutoff parameter $\rho_\text{min}$, and points that have $\rho_i < \rho_\text{min}$ are considered noise points.
A \defn{cluster center} is a point whose density is a local maximum within a cluster. 
Each cluster center corresponds to a separate cluster.
One way to pick cluster centers is using a parameter $\delta_\text{min}$, where
a point $x_i$ is considered a cluster center if $\delta_i > \delta_\text{min}$.

We use the \defn{work-span model}~\cite{JaJa92,CLRS}, a standard model of computation for analyzing shared-memory parallel algorithms.  
The \defn{work} $\mathcal{W}$ of an algorithm is the total number of operations executed by the algorithm, and the \defn{span} $\mathcal{S}$ is the length of the longest sequential dependence of the algorithm (it is also the parallel time complexity when there are an infinite number of processors).
We can bound the expected running time of an algorithm on $\mathcal{P}$ processors by $\mathcal{W}/\mathcal{P} + O(\mathcal{S})$ using a randomized work-stealing scheduler~\cite{Blumofe1999}.

\subsection{Relevant Techniques}\label{sec:relevant-technique}
\myparagraph{Graph-based Approximate Nearest Neighbor Search}
We use approximate nearest neighbor search (ANNS) algorithms in \framework. 
Graph-based ANNS algorithms can find approximate nearest neighbors in high dimensions efficiently and accurately compared to alternatives such as locality-sensitive hashing, inverted indices, and tree-based indices~\cite{wang2021comprehensive, ANNScaling, wanggraph}. 
These algorithms first construct a graph index on the input points, and later answer nearest neighbor queries by traversing the graph using a greedy search. Some popular methods include Vamana~\cite{jayaram2019diskann}, HNSW~\cite{Malkov2020ann}, HCNNG~\cite{munoz2019hcnng}, and PyNNDescent~\cite{pynndescent}. Manohar et al.~\cite{ANNScaling} provide parallel implementations for constructing these indices,
as well as a sequential implementation for running a single query. Multiple queries can be processed in parallel. 
We describe more graph-based ANNS methods in \Cref{sec:related}. 
Graph-based indices usually support any distance measure, while some indices~\cite{jayaram2019diskann, pynndescent} use heuristics that assume the triangle inequality holds. 

\myparagraph{ANNS on a Graph Index}
We use the function $G.$\algname{Find-Knn}$(x, k)$ to perform an ANNS on a graph $G$ for the point $x$, which returns the approximate \knn{s} of $x$. 
Most graph-based ANNS methods use a variant of a \textit{greedy (beam) search} (\Cref{alg:greedySearch}) to answer a \knn  query~\cite{ANNScaling}. For a query point $x$, the algorithm maintains a \defn{beam} $\mathcal{L}$ with size at most $L$ (the \defn{width} of the beam) as a set of candidates for the nearest neighbors of $x$.

Let $G.E_{\text{out}}(x)$ be the vertices incident to the edges going out from $x$ in $G$. We call these the \defn{out-neighbors} of $x$.
On each step, the algorithm pops the closest vertex
to $x$ from $\mathcal{L}$ (Line~\ref{alg:greedySearch:pop}), and processes it by adding all of its out-neighbors to the
beam (Line~\ref{alg:greedySearch:addneighbor}). 
The set $\mathcal{V}$ maintains all points that
have been processed
(Line~\ref{alg:greedySearch:updatev}). 
If $|\mathcal{L}|$ exceeds $L$, only the $L$ closest points to $x$ will be kept (Line~\ref{alg:greedySearch:prune}).
The algorithm stops when all vertices in the beam have been visited, as no new vertices can be explored (Line~\ref{alg:greedySearch:while}). 
The algorithm returns the $k$ closest points to $x$ from $\mathcal{L}$ and the visited point set $\mathcal{V}$ (Line~\ref{alg:greedySearch:return}).

In some cases,
it is possible that the algorithm traverses fewer than $k$ points for a query, and thus returns fewer than $k$ points. To solve this problem, options include using a brute force search or repeating the search from other starting points.

\begin{algorithm}[t]\small
\caption{Greedy Beam Search, modified from \cite{jayaram2019diskann}}
\label{alg:greedySearch}
\footnotesize
\begin{algorithmic}[1]
\Require Query point $x$, starting point set $S$, graph index $G$, beam width $L$, dissimilarity measure $D$, and integer $k$.
\State $\mathcal{V} \gets \emptyset$ \Comment{visited points}
\State $\mathcal{L} \gets S$ \Comment{points in the beam}
\While{$\mathcal{L} \setminus \mathcal{V} \neq \emptyset$} \label{alg:greedySearch:while}
    \State $p^* \gets \text{argmin}_{(q \in \mathcal{L} \setminus \mathcal{V})} D(x, q)$ \label{alg:greedySearch:pop}

    \State $\mathcal{L} \gets \mathcal{L} \cup G.E_{\text{out}}(p^*)$ \label{alg:greedySearch:addneighbor}
    \State $\mathcal{V} \gets \mathcal{V} \cup \{p^*\}$ \label{alg:greedySearch:updatev}
    \If{$|\mathcal{L}| > L$}
         keep only the $L$ closest points to $x$ in $\mathcal{L}$ \label{alg:greedySearch:prune}
    \EndIf
\EndWhile
\State \Return $k$ closest points to $x$ in $\mathcal{L} \cup \mathcal{V}$ \label{alg:greedySearch:return}
\end{algorithmic}
\end{algorithm}

\myparagraph{Parallel Primitives}
\algname{par-filter}($A$, $f$) takes as input a sequence of elements $A$ and a predicate $f$, and returns all elements $a\in A$ such that $f(a)$ is true. 
\algname{par-argmin}($A$, $f$) takes as input a sequence of elements $A$ and a function $f : A \to R$, and returns the element $a\in A$ that has the minimum $f(a)$. 
\algname{par-sum}($A$) takes as input a sequence of numbers $A$, and returns the sum of the numbers in $A$. \algname{par-filter}, \algname{par-argmin}, and \algname{par-sum} all take $O(n)$ work and $O(\log n)$ span. 
\algname{par-select}($A$, $k$) takes as input a sequence of elements $A$ and an integer $0 < k \leq |A|$, and returns the $k^\text{th}$ largest element in $A$. It takes $O(n)$ work and $O(\log n \log\log n)$ span~\cite{JaJa92}.
We use the implementations of these primitives from  ParlayLib~\cite{Blelloch20}.

A \defn{union-find} data structure maintains the set membership of elements and allows the sets to merge. Initially, each element is in its own set. A \algname{union}($a,b$) operation merges the sets containing $a$ and $b$ into the same set. A \algname{find}($a$) operation returns the membership of element $a$.
We use a concurrent union-find data structure~\cite{Jayanti21}, which supports operations in parallel.
Performing $m$ unions on a set of $n$ elements takes $O(m( \log(\frac{n}{m}+1) + \alpha(n,n)))$ work and $O(\log n)$ span ($\alpha$ denotes the inverse Ackermann function).

%% file: framework.tex
\section{\framework Framework}\label{sec:framework}

We present the \framework framework in \Cref{alg:framework}. To make our description of the framework more concrete, we will give an example of instantiating the framework in this section.  \Cref{sec:usage} will provide more examples and \Cref{sec:analysis} will provide the work and span analysis of \framework.

\input{framework_psuedo}

The input to \framework is a point set $P$, a positive integer $k$, a distance measure $D$, and three functions $F_{\text{density}}$, $F_{\text{noise}}$, and $F_{\text{center}}$ that indicate how the density, noise points, and center points are computed, respectively. In the pseudocode, $\rho$ is an array of densities of all points in $P$ and $\mathcal{N}$ is an array containing $k$-nearest neighbors for all points. $\lambda$ is an array containing dependent points. $c$ is an array containing the cluster IDs of all points and $c_i$ is the cluster ID of $x_i$.  
The framework has the following six steps.

\myparagraph{1. Construct Index} On Line~\ref{alg:framework:index}, we construct an index $G$, which can be any index that supports \knn search. For example, it can be
a \kdt, which is suitable for low-dimensional exact \knn search~\cite{Friedman77}, or a graph-based index for ANNS on high-dimensional data~\cite{ANNScaling, jayaram2019diskann, Malkov2020ann, munoz2019hcnng, pynndescent}. It can also be an empty data structure, which would lead to doing brute force searches to find the exact \knn{s}.
An example of a graph index corresponding to a point set is shown in \Cref{fig:data}.

\myparagraph{2. Compute $k$-nearest Neighbors} On Lines~\ref{alg:framework:knn-start}--\ref{alg:framework:knn}, we compute the \knn{s} of all points in parallel, using the index $G$. 
If we run the greedy search (\Cref{alg:greedySearch}) on the example in \Cref{fig:data} with $k=1, L=1$, and $S$ containing only the query point, we would find that the nearest neighbors of $a$, $b$, $c$, $d$, $e$, and $f$ are $c$, $c$, $b$, $f$, $d$, and $d$, respectively 
(here we assume that the graph index returns exact nearest neighbors).

\myparagraph{3. Compute Densities} On Lines~\ref{alg:framework:density-start}--\ref{alg:framework:density}, we compute the density for each point in parallel using $F_{\text{density}}$. 
An example density function is $\frac{1}{D(x_i, x_j)}$, where $x_j$ is the furthest neighbor from $x_i$ in $\mathcal{N}_i$~\cite{sddp}.
For this density function, the densities of the points in \Cref{fig:data} are $\rho_a = \frac{1}{\sqrt{2}}$, $\rho_b = 1$, $\rho_c = 1$, $\rho_d = \frac{1}{\sqrt{2}}$, $\rho_e = \frac{1}{2}$, and $\rho_f = \frac{1}{\sqrt{2}}$. 
The ranking of the densities from high to low (breaking ties by node ID) is $b, c, a, d, f, e$.

\myparagraph{4. Compute Dependent Points} On Line~\ref{alg:framework:dp}, we compute the dependent point of all points in parallel. 
The dependent points in our example are shown in \Cref{fig:dependent}.
We explain the details of how we compute the dependent points in \Cref{sec:dependent}.
As mentioned in \Cref{sec:intro}, the resulting tree from this step is a hierarchy of clusters (dendrogram), which can be returned if desired. To compute a specific clustering, the following two steps are needed.

\myparagraph{5. Compute Noise and Center Points} 
On Lines~\ref{alg:framework:noise}--\ref{alg:framework:center}, we compute the noise and center points using the input functions $F_{\text{noise}}$ and $F_{\text{center}}$. An example of $F_{\text{noise}}$ is $\texttt{par-filter}(P, x_i:\rho_i > \rho_{\min})$, where $\rho_{\min}$ is a user-defined parameter. Points whose densities are at most $\rho_{\min}$ are classified as noise points. 
An example of $F_{\text{center}}$ is $\texttt{par-filter}(P, x_i: D(x_i, \lambda_i) \geq \delta_{\min})$, where $\delta_{\min}$ is a user-defined parameter. Non-noise points whose distance are at least $\delta_{\min}$  from their dependent point are classified as center points.
In our example (\Cref{fig:cluster}), 
if we let $\rho_{\min} = \frac{1}{\sqrt{2}}$, then $e$ is a noise point. If we let $\delta_{\min} = 2.5$, then $b$ and $d$ are center points.

\myparagraph{6. Compute Clusters}
On Lines~\ref{alg:framework:inituf}--\ref{alg:framework:find}, we compute the clusters using a concurrent union-find data structure~\cite{Jayanti21}.
In parallel, for all points that are not noise points or center points, we merge them into the same cluster as their dependent point. 
This ensures that points (except noise points and center points) are in the same cluster as their dependent point.  
\Cref{fig:cluster} shows the clustering obtained on our example. 
Since $e$ is a noise point and $b$ and $d$ are center points, we skip processing their outgoing edge during the union step (Line~\ref{alg:framework:union} of \Cref{alg:framework}).

\input{examples/data}

%% file: framework_psuedo.tex
\begin{algorithm}[!t]
\vspace{-3pt}
 \footnotesize
\caption{\framework Framework}
 \begin{algorithmic}[1]
 \Require Point set $P$, integer $k>0$, distance measure $D$, $F_{\text{density}}$, $F_{\text{noise}}$, $F_{\text{center}}$
 \State $G$ = \algname{BuildIndex}($P$)\label{alg:framework:index}
 \ParFor{$i \in 1 \ldots n$}\label{alg:framework:knn-start}
 \State $\mathcal{N}_i \gets G$.\algname{Find-Knn}($x_i$, $k$) \Comment{find \knn{s}}\label{alg:framework:knn}
 \EndParFor
 \ParFor{$i \in 1 \ldots n$}\label{alg:framework:density-start}
 \State $\rho_i \gets F_{\text{density}}$($x_i$, $\mathcal{N}_i$) \Comment{compute densities}\label{alg:framework:density}
 \EndParFor
  \State $\lambda \gets $ \Call{ComputeDepPts}{$G$, $P$, $\rho$, $\mathcal{N}$, $D$} \label{alg:framework:dp}
 \State $P_{\text{noise}} \gets F_{\text{noise}}$($P$, $\rho$, $\lambda$, $\mathcal{N}$) \Comment{compute noise points}\label{alg:framework:noise}
 \State $P_{\text{center}} \gets F_{\text{center}}$($P \setminus P_{\text{noise}}$, $\rho$, $\lambda$, $\mathcal{N}$) \Comment{compute center points}\label{alg:framework:center}
 \State Initialize a union-find data structure $UF$ with size $n=|P|$ \label{alg:framework:inituf}
 \ParFor{$x_i \in P \setminus ( P_{\text{noise}} \cup P_{\text{center}})$}
 \State $UF$.\algname{Union}($i$, $\lambda_i$)\label{alg:framework:union}
 \EndParFor
 \ParFor{$i \in 1 \ldots n$}
 \State $c_i \gets$ $UF$.\algname{Find}($i$) \label{alg:framework:find}
 \EndParFor
 \State Return $c$\label{alg:framework:finish}
 \end{algorithmic}
 \label{alg:framework}
\end{algorithm}

\begin{algorithm}[!t]
\footnotesize
\caption{Dependent Point Computation} \label{alg:computedp}
\begin{algorithmic}[1]
\Function{DPBruteForce}{$x_i$, $\mathcal{N}_{\text{candidates}}$, $\rho$, $D$} \label{alg:depptsbruteforce}
    \State $\mathcal{N}_{\text{candidates}} \gets $ \Call{par-filter}{$\mathcal{N}_{\text{candidates}}$, $j:\rho_j > \rho_i$} \label{alg:depptsbruteforce:filter}
    \If{$\mathcal{N}_{\text{candidates}} = \emptyset$} \label{alg:depptsbruteforce:checkempty}
        \textbf{return} $\emptyset$ \label{alg:depptsbruteforce:returnempty}
    \EndIf
    \State $\lambda_i \gets $ \Call{par-argmin}{$\mathcal{N}_{\text{candidates}}$, $j:D(x_i, x_j)$} \label{alg:depptsbruteforce:argmin}
    \State \Return $\lambda_{i}$ \label{alg:depptsbruteforce:returnlambda}
\EndFunction

\Function{ComputeDepPts}{$G$, $P$, $\rho$, $\mathcal{N}$, $D$} \label{alg:computedeppts}
    \ParFor{$x_i \in P$} \label{alg:computedeppts:parforstart}
    \State $\lambda_i \gets $ \algname{DPBruteForce}($x_i$, $\mathcal{N}_i$, $\rho$, $D$) \label{alg:computedeppts:callbruteforce}
    \EndParFor \label{alg:computedeppts:parforend}
    \State $P_\text{unfinished} \gets $ \Call{par-filter}{$P$, $x_i:\lambda_i=\emptyset$} \label{alg:computedeppts:filterunfinished}
    \State $k^\text{dep} \gets L_d$  \Comment{$L_d$ is an integer parameter $> k$}\label{alg:computedeppts-opt:initL}
    \While{$|P_\text{unfinished}| > $ threshold} \label{alg:computedeppts:whilestart}
    \ParFor{$x_i \in P_\text{unfinished}$} \label{alg:computedeppts:parforunfinishedstart}
        \State $\mathcal{N}_{\text{candidates}} \gets$ $G$.\Call{FindKnn}{$i$, $k^\text{dep}$} \label{alg:computesingledeppts:findknn}
        \State $\lambda_i \gets $  \algname{DPBruteForce}($x_i$, $\mathcal{N}_{\text{candidates}}$, $\rho$, $D$) \label{alg:computesingledeppts:callbruteforce}
    \EndParFor \label{alg:computedeppts:parforunfinishedend}
    \State $k^\text{dep} \gets 2 \cdot k^\text{dep}$ \label{alg:computesingledeppts:updateL}   
    \State $P_\text{unfinished} \gets $ \Call{par-filter}{$P_\text{unfinished}$, $x_i:\lambda_i=\emptyset$}\label{alg:computedeppts:updateunfinished}
    \EndWhile \label{alg:computedeppts-opt:whileend}
    \ParFor{$x_i \in P_\text{unfinished}$} \label{alg:computedeppts:parforfinalstart}
    \State $\lambda_i \gets $ \algname{DPBruteForce}($x_i$, $P$, $\rho$, $D$) \label{alg:computedeppts:finalbruteforce}
    \EndParFor \label{alg:computedeppts:parforfinalend}
    \State \Return $\lambda$ \label{alg:computedeppts:returnlambda}
\EndFunction
\end{algorithmic}
\end{algorithm}

%% file: examples/data.tex
\noindent %
\begin{minipage}{.45\columnwidth}
    \centering
    \begin{tikzpicture}[scale=0.35]
      \def\points{(5,1)/a, (4,3)/b, (1,0)/e, (0,3)/f}
      \def\pointscd{(1,2)/d, (4,2)/c}

        \foreach \x in {0,1,...,4}
            \draw[thin,gray!30] (\x,-1) -- (\x,4);
        \foreach \y in {0,1,...,3}
            \draw[thin,gray!30] (-1,\y) -- (5,\y);

      \foreach \coord/\label in \points {
        \node[circle,fill,inner sep=2pt,label={[font=\Large]above right:\label}] (\label) at \coord {};
      }
        \foreach \coord/\label in \pointscd {
        \node[circle,fill,inner sep=2pt,label={[font=\Large]below left:\label}] (\label) at \coord {};
      }
    
  \draw[-] (f) -- node[midway,right]{} (d);
  \draw[-] (d) -- node[midway,right]{} (e);
  \draw[-] (f) -- node[midway,right]{} (c); 
  \draw[-] (c) -- node[midway,right]{} (b);
  \draw[-] (a) -- node[midway,right]{} (b);

    \end{tikzpicture}
        \vspace{-5pt}
    \captionof{figure}{Example dataset and a corresponding graph index.
    }
    \label{fig:data}
\end{minipage}
\hfill 
\begin{minipage}{.45\columnwidth} %
    \centering
    \begin{tikzpicture}[scale=0.35]
      \def\points{(5,1)/a, (4,3)/b, (1,0)/e, (0,3)/f}
      \def\pointscd{(1,2)/d, (4,2)/c}

      \foreach \x in {0,1,...,4}
          \draw[thin,gray!30] (\x,-1) -- (\x,4);
      \foreach \y in {0,1,...,3}
          \draw[thin,gray!30] (-1,\y) -- (5,\y);

    \foreach \coord/\label in \points {
      \node[circle,fill,inner sep=2pt,label={[font=\Large]above right:\label}] (\label) at \coord {};
    }
    \foreach \coord/\label in \pointscd {
        \node[circle,fill,inner sep=2pt,label={[font=\Large]below left:\label}] (\label) at \coord {};
      }
  
    \draw[-{Latex[width=6pt]}] (a) -- node[midway,right]{} (c);
    \draw[-{Latex[width=6pt]}] (c) -- node[midway,right]{} (b);
    \draw[-{Latex[width=6pt]}] (d) -- node[midway,right]{} (c);
    \draw[-{Latex[width=6pt]}] (f) -- node[midway,right]{} (d);
    \draw[-{Latex[width=6pt]}] (e) -- node[midway,right]{} (d);
    \end{tikzpicture}
        \vspace{-5pt}
    \captionof{figure}{Each point has an outgoing edge to its dependent point.}
    \label{fig:dependent}
\end{minipage}

\vspace{-10pt}

\begin{center}
\begin{minipage}{\columnwidth} %
    \centering
    \begin{tikzpicture}[scale=0.4]
      \def\points{(5,1)/a, (4,3)/b, (4,2)/c, (1,2)/d, (1,0)/e, (0,3)/f}

  \fill[blue!20, opacity=0.5] (4.5,2) circle (1.5cm); %

  \fill[blue!20, opacity=0.5] (0.5,2.25) circle (1.5cm); %

      \foreach \x in {0,1,...,4}
          \draw[thin,gray!30] (\x,-1) -- (\x,4);
      \foreach \y in {0,1,...,3}
          \draw[thin,gray!30] (-1,\y) -- (5,\y);

  \foreach \coord/\label in \points {
    \ifthenelse{\equal{\label}{b} \OR \equal{\label}{d}}{
      \ifthenelse{\equal{\label}{b}}{
        \node[diamond,fill=blue,inner sep=2pt,label={[font=\Large]above right:\label}] (\label) at \coord {};
      }{
        \node[diamond,fill=blue,inner sep=2pt,label={[font=\Large]below left:\label}] (\label) at \coord {};
      }
    }{
      \ifthenelse{\equal{\label}{e}}{
        \node[circle,draw,inner sep=2pt,label={[font=\Large]above right:\label}] (\label) at \coord {};
      }{
        \ifthenelse{\equal{\label}{c}}{
          \node[circle,fill,inner sep=2pt,label={[font=\Large]below left:\label}] (\label) at \coord {};
        }{
          \node[circle,fill,inner sep=2pt,label={[font=\Large]below left:\label}] (\label) at \coord {};
        }
      }
    }
  }

    \draw[-{Latex[width=6pt]}] (a) -- node[midway,right]{} (c);
    \draw[-{Latex[width=6pt]}] (c) -- node[midway,right]{} (b);
    \draw[-{Latex[width=6pt]}, dashed] (d) -- node[midway,right]{} (c);
    \draw[-{Latex[width=6pt]}] (f) -- node[midway,right]{} (d);
    \draw[-{Latex[width=6pt]}, dashed] (e) -- node[midway,right]{} (d);
    \end{tikzpicture}
    \vspace{-5pt}
    
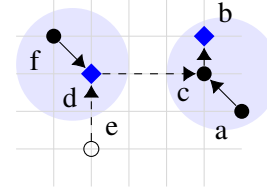
\captionof{figure}{Clustering result with $e$ as a noise point (white circle), and $b$ and $d$ as center points (blue diamonds). The dashed edges are ignored during the union step (Line~\ref{alg:framework:union} of \Cref{alg:framework}). The two blue circles are the two clusters found.
    }
    \label{fig:cluster}
\end{minipage}
\end{center}

%% file: optimization.tex
\subsection{Dependent Point Computation}\label{sec:dependent}

Our parallel algorithm for computing the dependent points (\Cref{alg:computedp}) takes as input the index $G$, the point set $P$, the array of densities $\rho$, the array of (approximate) \knn{s}  $\mathcal{N}$, and the distance measure $D$. 

\algname{DPBruteForce} 
is a helper function (Lines~\ref{alg:depptsbruteforce}--\ref{alg:depptsbruteforce:returnlambda}) that searches for the nearest neighbor of $x_i$ with density higher than $\rho_i$ among $\mathcal{N}_{\text{candidates}}$ using brute force. It returns $\emptyset$ if no points in $\mathcal{N}_{\text{candidates}}$ have a higher density than $\rho_i$.

On Lines~\ref{alg:computedeppts:parforstart}--\ref{alg:computedeppts:parforend}, we first search within the \knn{s} of each point to find its dependent point. This optimization is also used in several other works~\cite{sddp, sun2021density, chen2020fast}. On Line~\ref{alg:computedeppts:filterunfinished}, we obtain the set of points $P_\text{unfinished}$ that have not found their dependent points.
Line~\ref{alg:computedeppts-opt:initL} initializes $k^\text{dep}$ to $L_d$.

$L_d$ and \texttt{threshold} are parameters used for our performance optimizations. We defer a discussion of these parameters to \Cref{sec:optimization}, and ignore their effect here by setting $L_d$ to be $2k$ and \texttt{threshold} to be 0 (this causes Lines~\ref{alg:computedeppts:parforfinalstart}--\ref{alg:computedeppts:parforfinalend} to have no effect, since $P_\text{unfinished}$ will be empty at that point).

The while-loop on Line~\ref{alg:computedeppts:whilestart} terminates when all points have found their dependent point.
On Lines~\ref{alg:computedeppts:parforunfinishedstart}--\ref{alg:computedeppts:parforunfinishedend}, we compute the dependent point for points in $P_\text{unfinished}$. 
If the index is designed for approximate \knn search, we guarantee that the dependent point has a higher density, but it might not be the closest among points with higher densities.
Note that on Line~\ref{alg:computedeppts:parforunfinishedstart}, we can skip the point with maximum density, since we know that it does not have a dependent point.
On Lines~\ref{alg:computesingledeppts:findknn}--\ref{alg:computesingledeppts:callbruteforce}, for each point, we find $k^\text{dep}$ neighbors of $x_i$ on each round, and if any of the neighbors have a higher density than $x_i$, we can return the closest such neighbor as the dependent point. 
We then double $k^\text{dep}_i$ for the next round (Line~\ref{alg:computesingledeppts:updateL}). A similar doubling optimization is used in \cite{chen2020fast}, but with a cover tree. Furthermore, their algorithm is sequential.
On Line~\ref{alg:computedeppts:updateunfinished}, we compute the set of points $P_\text{unfinished}$ that have not found their dependent point.

\myparagraph{Example} On the dataset from \Cref{fig:data}, points $a$, $c$, $e$, and $f$ would find their dependent point within their \knn ($k=1$) on  Lines~\ref{alg:computedeppts:parforstart}--\ref{alg:computedeppts:parforend} because their nearest neighbor has higher density than themselves. $b$ is the point with maximum density and is skipped. 
For the remaining point $d$, on the first round we have $k^\text{dep}=2$, and so $\mathcal{N}_{\text{candidates}}=\{e, f\}$. This does not contain any point with a higher density than $d$, and so we double $K^\text{dep}=2$ and try again. On the second round, $k^\text{dep}=4$, and so $\mathcal{N}_{\text{candidates}}=\{b, c, e, f\}$, which contains $d$'s dependent point $c$.

\subsection{Performance Optimizations}\label{sec:optimization}

\subsubsection*{Dependent Point Finding}
Now we explain the two integer parameters $L_d$ and \texttt{threshold}. 
The while-loop on Line~\ref{alg:computedeppts:whilestart}
checks if $|P_\text{unfinished}| > $ \texttt{threshold}, and when that is no longer true, we do a brute force \knn computation for the remaining points in $P_\text{unfinished}$ on Lines~\ref{alg:computedeppts:parforfinalstart}--\ref{alg:computedeppts:parforfinalend}.
This optimization is useful because for the points with relatively high density, it can be challenging for the index to find a dependent point (as most neighbors have lower density than them), and for these last few points it is faster to just do a brute force search than continue to double $k^\text{dep}$. Furthermore, when few points are remaining, there is less parallelism available when calling \algname{FindKNN}, each of which is sequential, compared to the brute force search, which is highly parallel.  
In our experiments, we set $\texttt{threshold}=300$, which we found to work well. 

$L_d$ is a tunable parameter that is $> k$ (Line~\ref{alg:computedeppts-opt:initL}) and indicates the initial number of nearest neighbors to search for to find a dependent point (Line~\ref{alg:computesingledeppts:findknn}). 
A larger value of $L_d$ leads to fewer iterations. However, points that require fewer than $L_d$ nearest neighbors to find a dependent point will do some extra work (as they search for more nearest neighbors than necessary).
On the other hand, points that require at least $L_d$  nearest neighbors to find a dependent point will do less work overall (they do not need to waste work on the initial rounds where they would not find a dependent point anyway).

\myparagraph{Vamana Graph Construction}
Vamana~\cite{jayaram2019diskann, ANNScaling} is one of the graph-based indices that we use for ANNS. 
Its parallel construction algorithm~\cite{ANNScaling}  builds the graph by running greedy search (from \Cref{alg:greedySearch}) on each point $x_i$ (in batches), and then adds edges from $x_i$ to points visited during the search ($\mathcal{V}$). It requires a degree bound parameter $R$, such that in the constructed graph each vertex has at most $R$ out-neighbors. If adding edges between $x_i$ and $\mathcal{V}$ causes a vertex's degree to exceed $R$, a pruning procedure is called to iteratively select at most $R$ out-neighbors. 
The pruning algorithm also has a parameter $\alpha \geq 1$ 
that controls how aggressive the pruning is; a higher $\alpha$ corresponds to more aggressive pruning, which can lead to less than $R$ neighbors being selected. 
Intuitively, this heuristic prunes the long edge of a triangle, with a slack of $\alpha$. 
The details of the pruning algorithm can be found in~\cite{jayaram2019diskann}.

The original Vamana graph construction algorithm~\cite{jayaram2019diskann, ANNScaling} starts the greedy search from a single point, which is the medoid of $P$.
Starting from a single point can make the algorithm require a high degree bound and beam width to achieve good results on clustered data because a search can be trapped within the cluster that the medoid is in. 
Instead of using a large degree bound and beam width, which degrades performance, we use an optimization where we randomly sample a set of starting points for the Vamana graph construction algorithm instead of starting from the medoid alone. 
This heuristic is also explored in \cite{lin2019comparative}.

%% file: usage.tex
\section{Usage of \framework}\label{sec:usage}
\framework allows users to plug in functions that can be combined to obtain new clustering algorithms. In this section, we describe several functions and provide their work and span bounds.

\subsection{Indices}
Here we describe several approaches for building indices for \knn search.
Let the work and span of constructing $G$ be $\mathcal{W}_c$ and $\mathcal{S}_c$, respectively.

\myparagraph{Brute Force}
The brute force approach does not use an index at all.
When searching for the exact \knn{s} of $x_i$, it uses a \algname{par-select} to find the $k^{\text{th}}$ smallest distance to $x_i$, and a \algname{par-filter} to filter for the points with smaller distances to $x_i$.
In this case, $\mathcal{W}_c$ and $\mathcal{S}_c$ are $O(1)$, while $\mathcal{W}_{nn}$ and $\mathcal{S}_{nn}$ are $O(n)$ and $O(\log n \log\log n)$, respectively.

\myparagraph{Tree Indices}
Another option is to use a tree index, such as a \kdt or a cover tree~\cite{chen2020fast}. For a parallel \kdt, $\mathcal{W}_c = O(n \log n)$ and $\mathcal{S}_c = O(\log n \log\log n )$~\cite{wang2022esa}. 
A parallel cover tree can be constructed in $O(n \log n)$ expected work and $O(\log^3 n \log\log n)$ span with high probability~\cite{gu2022parallel}.
A \knn search in a \kdt takes $O(n)$ work and $O(\log n)$ span. A \knn search in a cover tree takes $O(c^7(k + c^3) \log k \log \Delta)$ expected work and span~\cite{gu2022parallel, elkin2023new, elkin2022counterexamples}, where $c$ is the expansion constant of $P$ and $\Delta$ is the aspect ratio of $P$. However, note that these tree indices usually suffer from the curse of dimensionality and do not perform well on high-dimensional datasets.

\myparagraph{Graph Indices}
Graph-based ANNS algorithms have been shown to be efficient and accurate in finding approximate nearest neighbors in high dimensions~\cite{wang2021comprehensive, ANNScaling, wanggraph}.
Our framework includes three parallel graph indices from the ParlayANN library~\cite{ANNScaling}: Vamana~\cite{jayaram2019diskann}, HCNNG~\cite{munoz2019hcnng}, and PyNNDescent~\cite{pynndescent}. Similar to Vamana, HCNNG also uses the parameter $\alpha$ to prune edges. \algname{HCNNG} and \algname{PyNNDescent} also accept a $num\_repeats$ argument, which represents how many times they will independently repeat the construction process before merging the results together.

When the number of returned neighbors is less than $k$, we use the brute force method to find the exact \knn{s}. While these graph indices have been shown to work well in practice, there are only a few works that theoretically analyze their performance~\cite{navarro2002searching, prokhorenkova2020graph, shrivastava2023theoretical, laarhoven2018graph, indyk2023worstcase}. 
Indyk and Xu~\cite{indyk2023worstcase} show that Vamana construction takes 
$\mathcal{W}_c=O(n^3)$ 
work. In practice, we find that the work is usually much lower.
Using the batch insertion method \cite{ANNScaling}, which inserts points in batches of doubling size, Vamana construction takes 
$\mathcal{S}_c = O(n^2 \log n) $ span.\footnote{The batch insertion method in \cite{ANNScaling} sets a batch size upper bound of $0.02n$, which does not affect the bounds, as there will only be a constant number ($<50$) more batches after the upper bound is reached.}

\subsection{Density, Center, and Noise Functions }
\label{sec:density}
Here, we describe a subset of the density, center, and noise functions ($F_{\text{density}}$, $F_{\text{center}}$, and $F_{\text{noise}}$) that we implement in \framework. 
We describe other functions we implement in 
\iffull
\Cref{sec:appendix-functions}.
\fi

\myparagraph{\texttt{kth} Density Function} 
The density of $x_i$ is $\rho_i = \frac{1}{D(x_i, x_j)}$ where $x_j$ is the furthest neighbor from $x_i$ in $\mathcal{N}_i$, i.e., the distance to the exact or approximate $k^{\text{th}}$ nearest neighbor of $x_i$~\cite{sddp, chen2020fast}.  
Each density computation is $O(k)$ work and $O(\log k)$ span to find the furthest neighbor in $\mathcal{N}_i$. 

The density can also be normalized~\cite{hou2019enhancing}. The normalized density (\textbf{\texttt{normalized}}) is $\rho'_i = \frac{\rho_i k }{\sum_{j \in \mathcal{N}_i} \rho_j}$. Intuitively, this function normalizes a point's density with an average of the densities of its neighbors.
Each normalization takes an extra $O(k)$ work and $O(\log k)$ span.

\myparagraph{Threshold Center Function} Recall from \Cref{sec:prelim} that $\delta_i =  D(x_i,\lambda_i)$ is the dependent distance of $x_i$. $F_{\text{center}}$ obtains the center points by selecting the points whose distance to their dependent point is greater than $\delta_{\min}$, a user-defined parameter. This can be implemented with a $\texttt{par-filter}$, whose work and span are $O(n)$ and $O(\log n)$, respectively. This method is used in \cite{Amagata21, yaohui2017adaptive, amagata2023efficient}. 

\myparagraph{Product Center Function} 
This method takes as input $n_{c}$, a user-defined parameter that specifies how many clusters to output.
We compute the product $\delta_i \times \rho_i$ for all points $x_i$. 
The $n_{c}$ points with the largest products are the center points. 
This function can be implemented with a \algname{par-select} to find the $n_{c}^{\text{th}}$ largest product $t$, and then a \algname{par-filter} to filter out the points with product less than $t$. The work and span are $O(n)$ and $O(\log n \log\log n)$, respectively.
This method is used in \cite{jiang2020adaptive, Rodriguez14, hou2019enhancing, liu2018shared}.

\myparagraph{Noise Function}
We implement a noise function $F_{\text{noise}}$, which returns the points $x_i$ with density $\rho_i < \rho_{\min}$. These points are then ignored in the remainder of the algorithm.
This can be implemented using a parallel filter with $O(n)$ work and $O(\log n)$ span. 
This noise function is used by \cite{Amagata21, Rodriguez14, amagata2023efficient}.

%% file: analysis.tex
\section{Analysis of \framework}\label{sec:analysis}

\subsection{Work and Span Analysis}

The work and span of \framework (\Cref{alg:framework}) depend on the specific index construction algorithm and functions $F_{\text{density}}$, $F_{\text{noise}}$, and $F_{\text{center}}$. Here, we choose the functions that give the best performance in our experiments (\texttt{kth} density, product center, and default noise functions).

We first analyze the work and span of computing dependent points as shown in Algorithm~\ref{alg:computedp} (this is called on Line~\ref{alg:framework:dp} of \Cref{alg:framework}).
Let $n_{\text{can}}=|\mathcal{N}_{\text{candidates}}|$.
Lines~\ref{alg:depptsbruteforce}--\ref{alg:depptsbruteforce:returnlambda} take $O(n_{\text{can}})$ work and $O(\log n_{\text{can}})$ span. Thus, Lines~\ref{alg:computedeppts:parforstart}--\ref{alg:computedeppts:parforend} take $O(nk)$ work and $O(\log k)$ span, because $|\mathcal{N}_i|=k$ and $|P|=n$.
Line~\ref{alg:computedeppts:filterunfinished} takes $O(n)$ work and $O(\log n)$ span.  

On Lines~\ref{alg:computedeppts:whilestart}--\ref{alg:computedeppts-opt:whileend}, for each point, we call \algname{G.FindKnn} $O(\log n)$ times since we double $k^{\text{dep}}$ after each round. 
Let the work and span of finding the $k$ nearest neighbors using $G$ be $\mathcal{W}_{nn}(k)$ and $\mathcal{S}_{nn}(k)$, respectively.
Let $\mathcal{W}_{nn} = \sum_{j=0}^{O(\log n)} \mathcal{W}_{nn}(2^j)$ and $\mathcal{S}_{nn} = \sum_{j=0}^{O(\log n)} \mathcal{S}_{nn}(2^j)$.
The filter on Line~\ref{alg:computedeppts:updateunfinished} takes $O(n\log n)$ work and $O(\log^2 n)$ span across $O(\log n)$ rounds.
The total work and span across all rounds is $O(n \mathcal{W}_{nn})$ and $O(\mathcal{S}_{nn} + \log^2 n)$.
The brute force computation on Lines~\ref{alg:computedeppts:parforfinalstart}--\ref{alg:computedeppts:parforfinalend} takes $O(n)$ work and $O(\log n)$ span, as $O(1)$ points remain after the loop on Lines~\ref{alg:computedeppts:whilestart}--\ref{alg:computedeppts-opt:whileend}.

Thus, the work and span of \Cref{alg:computedp} are $O(n \mathcal{W}_{nn})$ and $O(\mathcal{S}_{nn} + \log^2 n)$, respectively.

We now analyze the remaining steps of \Cref{alg:framework}.
Lines~\ref{alg:framework:knn-start}--\ref{alg:framework:knn}   compute the \knn{s} of all points, which takes $O(n\mathcal{W}_{nn}(k))$ work and $O(n\mathcal{S}_{nn}(k))$ span. 
Lines~\ref{alg:framework:density-start}--\ref{alg:framework:density} compute the densities of all points. Using the \texttt{kth} density function, this takes $O(nk)$ work and $O(\log k)$ span.
Lines~\ref{alg:framework:noise}--\ref{alg:framework:center} using the product center and default noise functions take $O(n)$ work and $O(\log n\log\log n)$ span.
The union-find operations on Lines~\ref{alg:framework:inituf}--\ref{alg:framework:find} take $O(n \alpha(n,n))$ work and $O(\log n)$ span.

The following theorem gives the overall work and span.

\begin{theorem}\label{thm:framework}
    The work and span of \framework using the $k^\text{th}$ density, product center, and the default noise functions are $O(\mathcal{W}_c + n \mathcal{W}_{nn})$ and $O(\mathcal{S}_c + \mathcal{S}_{nn} + \log^2 n)$, respectively.
\end{theorem}

\subsection{Approximation Analysis}\label{sec:approximation-analysis}
In this section, we give a brief analysis of the approximation guarantees of \framework. 
Proofs and more detailed analyses can be found in 
\iffull
\Cref{sec:appendix-approximation-analysis}.
\fi
Our analysis of the density approximation is based on the \texttt{kth} density function described above. Our analysis of the approximate dependent point computation is based on the threshold center 
function described above.

\myparagraph{Density Estimation}
Assuming some guarantee in approximate $k$-nearest neighbor search, we can show that the density peaks of the exact algorithm that do not \textit{conflict} with other points will remain density peaks. A conflict occurs when the density ranges of two points overlap. The density range of a point bounds the approximate density value of the point. 

\begin{restatable}{mylemma}{densitylemma}
Consider the threshold center function, which obtains the center points by selecting
the points whose distance to their dependent point is greater than $\delta_{\min}$. If the density interval of a point does not conflict with any other interval and it is a true density peak, then it is still a density peak in \framework given the same threshold $\delta_{\min}$.
\end{restatable}

Note that there may be additional density peaks returned by the approximate algorithm, but the true density peaks in the exact algorithm are guaranteed to still be density peaks.

\myparagraph{Dependent Point Estimation}
Now we analyze the approximate dependent point found by \Cref{alg:computedeppts}. 
The following lemma guarantees that the approximate dependent points returned by our algorithm are not too much further than the true dependent points.
Let $d_j$ be the distance to the true $j^\text{th}$ nearest neighbor from query point $q$. As far as we know, other approximate DPC methods~\cite{amagata2022scalable, Amagata21, gong2017clustering} do not provide approximation bound on approximate dependent point search.

\begin{restatable}{mylemma}{dependentpointlemma}
\label{lemma:dependent-point}
Suppose we find the approximate dependent point among the $\beta k$-approximate nearest neighbor, for $\beta \geq 1$. The approximate dependent point is at most $c^2 \frac{d_{\beta k}}{d_k}$ further from the exact dependent point given the same densities for some constant $c \geq 1$.
\end{restatable}

In \Cref{alg:computedeppts}, we use $\beta=2$
 for \Cref{lemma:dependent-point}, since we double the number of nearest neighbors to find until we have found a dependent point.

%% file: experiment.tex
\section{Experiments}

\begin{table}[t]
\footnotesize
\centering
\setlength{\tabcolsep}{3pt}
\vspace{-3pt}
 \begin{tabular}{l c c c c c} 
 \toprule
 Name & $n$ & $d$ & Description & \# Clusters\\ %
 \midrule
 \datasetname{gaussian} & $10^5$ to $10^8$ & 128 & Standard benchmark  & 10 to 10000 \\
 \datasetname{MNIST} & 70,000 & 784 & Raw images & 10\\
 \datasetname{ImageNet} & 1,281,167& 1024 & Image embeddings & 1000 \\
  \datasetname{birds} & 84,635 & 1024 & Image embeddings & 525 \\
\datasetname{reddit} & 420,464 & 1024 & Text embeddings &50 \\
\datasetname{arxiv} & 732,723 & 1024 & Text embeddings &180 \\
 \bottomrule
\end{tabular}
\caption{\label{tab:datasets}Our datasets, along with their sizes ($n$), their dimensionality ($d$), and the number of ground truth clusters.}
\end{table}

\subsection{Experimental Setup}\label{sec:exp-setup}
\subsubsection*{Computational Environment}
We use \textit{c2-standard-60} instances on the Google Cloud Platform. These are 30-core machines with two-way hyper-threading with Intel 3.1 GHz Cascade Lake processors that can reach a max turbo clock-speed of 3.8 GHz. The instances have two non-uniform memory access (NUMA) nodes, each with 15 cores. Except for the experiments studying scalability with respect to the number of threads, we use all 60 hyper-threads for our experiments.

\myparagraph{Datasets}
We use a variety of real-world and artificial datasets, summarized in \Cref{tab:datasets} and described below.

\begin{list}{\textbullet}{%
    \setlength{\leftmargin}{0.5em}
    \setlength{\itemindent}{0em}
    \setlength{\labelwidth}{\itemindent}
    \setlength{\labelsep}{0.5em}
    \setlength{\listparindent}{1em}
    \setlength{\itemsep}{0em}
    \setlength{\parsep}{0em}
    \setlength{\topsep}{0em}
    \setlength{\partopsep}{0em}
}   
    \item \datasetname{gaussian} is a synthetic mixture of datasets generated from a Gaussian distribution. To generate a \datasetname{gaussian} dataset of dimension $d = 128$ with size $n$ and $c$ clusters, we first sample $c$ centers $x_i$ uniformly from $[0, 1]^d$, and then sample $n / c$ points from a Gaussian centered at each $x_i$ with variance $0.05$.

    \item \datasetname{MNIST}~\cite{deng2012mnist} is a standard dataset that consists of $28 \times 28$ dimensional images of grayscale digits between $0$ and $9$. The $i^\text{th}$ cluster corresponds to all occurrences of digit $i$.

    \item \datasetname{ImageNet}~\cite{imagenet} is a standard image classification benchmark with more than one million images, each of size $224 \times 224 \times 3$. The images are from $1000$ classes of everyday objects. Unlike for MNIST, we do not cluster the raw \datasetname{ImageNet} images, but instead first pass each image through  ConvNet~\cite{liu2022convnet} to get an embedding. Each ground truth cluster contains the embeddings corresponding to a single image class from the original \datasetname{ImageNet} dataset.

    \item \datasetname{birds}~\cite{birds} is a dataset that contains images of $525$ species of birds. The images have the same number of dimensions as \datasetname{ImageNet}, and we pass it through the same ConvNet model to obtain an embedding dataset. The ground truth clusters are the $525$ species of birds. This dataset is is out of distribution for the original ConvNet model. 

    \item \datasetname{reddit} and \datasetname{arxiv} are text embedding datasets studied in the recent Massive Text Embedding Benchmark  (MTEB) work~\cite{muennighoff2022mteb}. We restrict our attention to embeddings from the best model on the current MTEB leaderboard, GTE-large~\cite{li2023towards}. We also restrict our attention to the two largest datasets from MTEB, \datasetname{reddit}, where the goal is to cluster embeddings corresponding to post titles into subreddits, and \datasetname{arxiv}, where the goal is to cluster embeddings corresponding to paper titles into topic categories.
\end{list}

\begin{table}[!t]
\vspace{-3pt}
\footnotesize
\centering
\begin{tabular}{c c c c c }
\toprule
Dataset & $L$ & $L_d$ & $R$ & $k$\\ \hline
\datasetname{MNIST} & 32 & 32 & 32 & 16\\ 
\datasetname{ImageNet} & 128 & 128 & 128 & 16\\
\datasetname{reddit, arxiv} & 64 & 64 & 64 & 16\\
\datasetname{gaussian, birds} & 32 & 32 & 32& 16 \\ 
\hline
\end{tabular}
\caption{Default parameters used for datasets.
}
\label{tab:params}
\end{table}

\myparagraph{Algorithms}
We implement our algorithms using the ParlayLib~\cite{Blelloch20} and ParlayANN~\cite{ANNScaling} libraries.  
We use C++ for all implementations, and the gcc compiler with the \texttt{-O3} flag to compile the code. We also provide Python bindings for \framework.
We evaluate the following algorithms. 
\begin{list}{\textbullet}{%
    \setlength{\leftmargin}{0.5em}
    \setlength{\itemindent}{0em}
    \setlength{\labelwidth}{\itemindent}
    \setlength{\labelsep}{0.5em}
    \setlength{\listparindent}{1em}
    \setlength{\itemsep}{0em}
    \setlength{\parsep}{0em}
    \setlength{\topsep}{0em}
    \setlength{\partopsep}{0em}
}
    \item \framework: Our framework described in \Cref{sec:framework} with the different density functions described in \Cref{sec:usage}. Unless specified otherwise, we use the \texttt{kth} density function without normalization with $k = 16$, the \algname{Vamana} graph index with $\alpha = 1.1$, and the product center function with $n_{c}$ set to the number of ground truth clusters, and the default noise function.
    In \Cref{tab:params}, we give the rest of the default parameters that we used for each dataset.
    \item \algname{fastdp}~\cite{sieranoja2019fast}: A single-threaded approximate DPC algorithm that also uses graph-based ANNS to estimate densities. 
    \item \algname{$k$-means}: The FAISS~\cite{johnson2019billion} implementation of $k$-means, an extremely efficient $k$-means implementation. It is parallelized by using parallel $k$-nearest neighbor search.
    The $k$-means algorithm takes in $k$, the number of clusters, $niter$, the number of iterations, and $nredo$, the number of times to retry and choose the best clustering. Unless specified otherwise, the number of clusters used in $k$-means is the number of clusters in the ground truth clustering. 
    \item \algname{BruteForce}: An instantiation of \framework, where we use a naive parallel brute force approach for every step. This method takes $O(n^2)$ work. It also 
    first searches within the \knn{s} to find the dependent point.
    We refer to the result of \algname{BruteForce} as the "exact DPC" result.
    \item \algname{DBSCAN}: A density-based clustering algorithm for low-dimensional data~\cite{Ester96,schubert2017dbscan}. 
    We use the implementation in the Intel Extension for Scikit-learn~\cite{scikit-learn} for high-dimensional datasets, which is implemented in C++ and parallelized with parallel nearest neighbor search.
    We also tried Wang et al.'s~\cite{wang2019dbscan} parallel implementation, which is optimized for low-dimensional data, and found it slower than Scikit-learn on high-dimensional data.
    DBSCAN has two parameters $\epsilon$ and $min\_pts$: $\epsilon$ defines the maximum distance between two points to be considered neighbors. $min\_pts$ specifies the minimum number of points required to form a dense region (core point), which triggers the formation of a cluster. 
    
\end{list}

We also tried a parallel exact DPC algorithm that uses a priority search \kdt-based dependent point finding algorithm that was designed for low dimensions~\cite{dpc}. We changed the first step of \cite{dpc} from a range search to a \knn search to match our framework. On \datasetname{MNIST}, their algorithm takes 280s 
on our 30-core machine, which is $320$ times slower than \framework. This method is prohibitively slow because {\kdt}s suffer from the curse of dimensionality, where performance in high dimensions degrades to no better than a linear search~\cite{weber1998quantitative}. We thus do not further compare against this method.

\myparagraph{Evaluation}
We evaluate clustering quality using the Adjusted Rand Index (ARI)~\cite{Hubert1985}, homogeneity, and completeness~\cite{rosenberg2007v}.
Consider our clustering $\mathcal{C}$ and the ground-truth or exact clustering $\mathcal{T}$. Intuitively, ARI evaluates how similar $\mathcal{C}$ and $\mathcal{T}$ are. Homogeneity measures if each cluster in $\mathcal{C}$ contains members from the same class in $\mathcal{T}$. Completeness measures whether all members in $\mathcal{T}$ of a given class are in the same cluster in $\mathcal{C}$. 

Let $n_{ij}$ be the number of objects in the ground truth cluster $i$ and the cluster $j$ generated by the algorithm, $n_{i*}$ be $\sum_j n_{ij}$, $n_{*j}$ be $\sum_i n_{ij}$, 
and $n$ be $\sum_i n_{i*}$. The ARI is computed as 
$\frac{\sum_{i,j} {n_{ij}\choose 2} - [\sum_i {n_{i*}\choose 2} \sum_j {n_{*j}\choose 2} ] / {n\choose 2}}{\frac{1}{2}[\sum_i {n_{i*}\choose 2} +\sum_j {n_{*j}\choose 2} ] - [\sum_i {n_{i*}\choose 2} \sum_j {n_{*j}\choose 2} ]  /   {n \choose 2}}$.
The ARI score is $1$ for a perfect match, and its expected value is $0$ for random assignments.

The formulas for homogeneity and completeness of clusters are defined as follows:
    $\text{homogeneity} = 1 - \frac{H(\mathcal{C}|\mathcal{T})}{H(\mathcal{C})}$;  
    $\text{completeness} = 1 - \frac{H(\mathcal{T}|\mathcal{C})}{H(\mathcal{T})}$.
\( H(\mathcal{C}|\mathcal{T}) \) is the conditional entropy of the class distribution given the cluster assignment, \( H(\mathcal{C}) \) is the entropy of the class distribution, \( H(\mathcal{T}|\mathcal{C}) \) is the conditional entropy of the cluster distribution given the class, and \( H(\mathcal{T}) \) is the entropy of the cluster distribution.
For example, consider a ground-truth clustering $\mathcal{T}$ where all classes have the same number of points. If $\mathcal{C}$ assigns every point to its own cluster of size $1$, it has homogeneity score $1$ and a low completeness score when $n_c \ll n$. 
If $\mathcal{C}$ assigns all points to a single cluster, it has completeness score $1$ and homogeneity score $0$.

\begin{figure}[t]
 \centering
 \vspace{-1pt}
 \includegraphics[width=0.9\columnwidth]{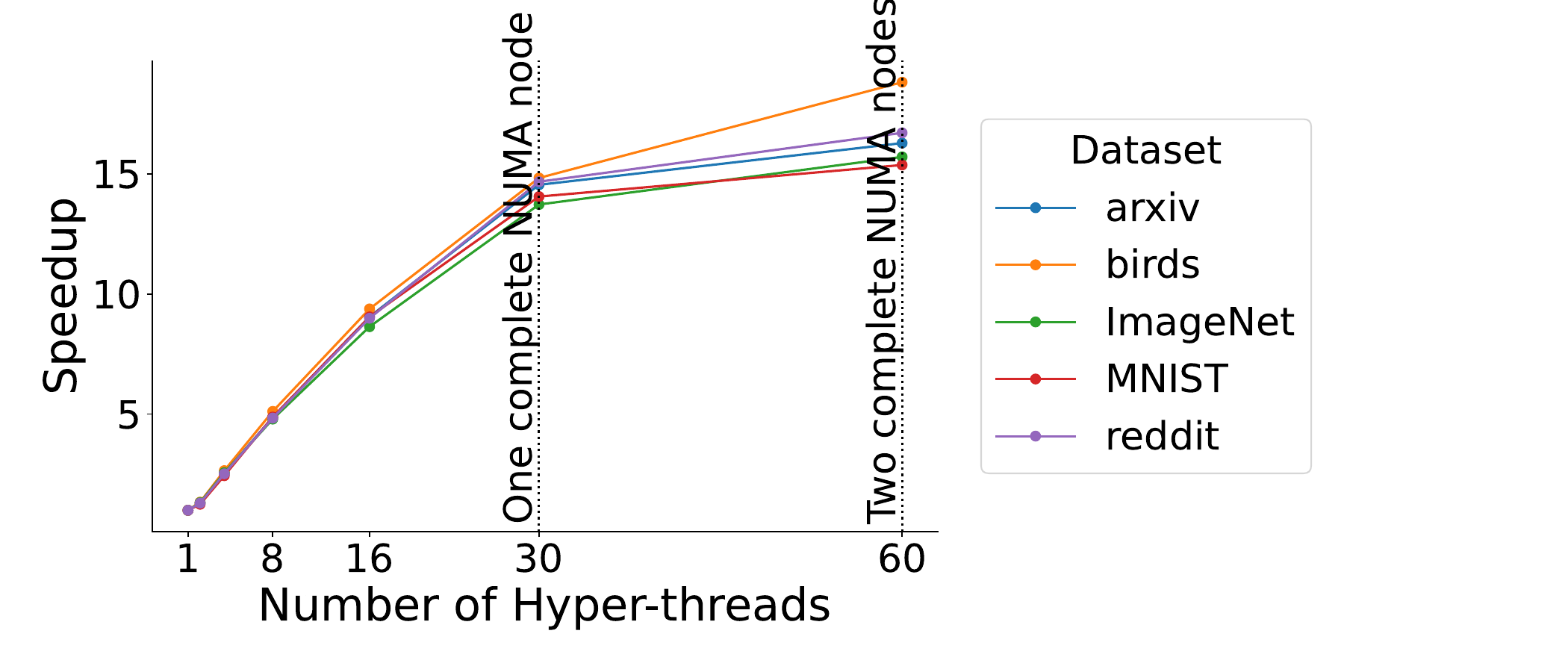}
 \vspace{-5pt}
  \caption{Self-relative parallel speedup across different numbers of hyper-threads. }
  \label{fig:dpc-scalability-thread}
\end{figure}

\subsection{Scalability}
Figure \ref{fig:dpc-scalability-thread} shows the parallel scalability of \framework on our larger datasets. 
\framework achieves an average of $14.36$x self-relative speedup on one NUMA node with 30 hyper-threads and an average of $16.57$x self-relative speedup on two NUMA nodes with 60 hyper-threads. 

We also study the runtime of \framework as we increase the size of the synthetic \datasetname{gaussian} dataset and vary the number of clusters
\iffull
 between 10 to 10,000 (Figure~\ref{fig:dpc-scalability-n}). We use a linear fit on the logarithm of runtime and $ \log n$ to obtain the slopes of the lines in \Cref{fig:dpc-scalability-n}. The slope $s$ reflects the exponent in the growth of runtime with respect to data size. 
\fi
We find that the slope ranges from 1.12--1.2 depending on the number of output clusters, and thus experimentally the runtime grows approximately as $O(n^{1.2})$ for this dataset. 
This shows that \framework has good scalability with respect to $n$.

\iffull
\begin{figure*}[t]
  \centering \includegraphics[width=0.9\textwidth]{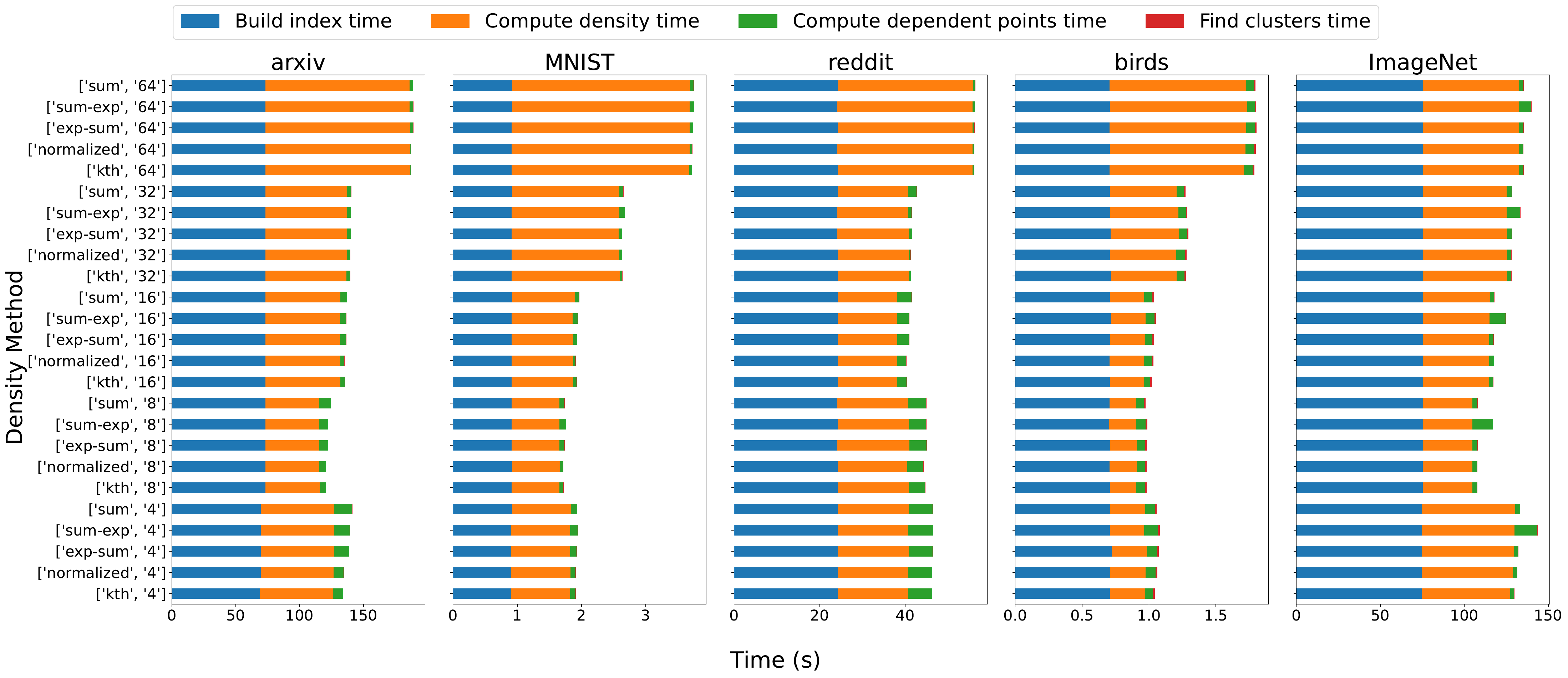}
  \caption{Runtime decomposition of \framework with different density functions and values of $k$.}
  \label{fig:decomposition}
  \vspace{0.3cm}
\end{figure*}
\fi

\subsection{Runtime Decomposition}
We present the runtime decomposition of \framework on each dataset with all density methods and all values of $k$ in 
\iffull
\Cref{fig:decomposition} and \Cref{tab:small_runtime}.
\fi
The bottleneck of the runtime is the index construction time and the \knn time when computing densities. 
When $k$ is larger, the \knn search time for density computation is longer, as expected.
Computing clusters with union-find is fast because this step has low work, as discussed in \Cref{sec:usage}.
The dependent point computation time is much shorter than the density computation because
the dependent point for some points can be obtained from the \knn{s} (Lines~\ref{alg:computedeppts:parforstart}--\ref{alg:computedeppts:parforend} in \Cref{alg:computedeppts}), so we do not need to run nearest neighbor searches for these points. Additionally, even when the dependent point is not in the \knn{s}, our doubling technique finds a dependent point in the first few rounds for most points, thereby usually avoiding an expensive exhaustive search.

\begin{figure*}[t]
\vspace{-5pt}
  \centering \includegraphics[width=\textwidth]{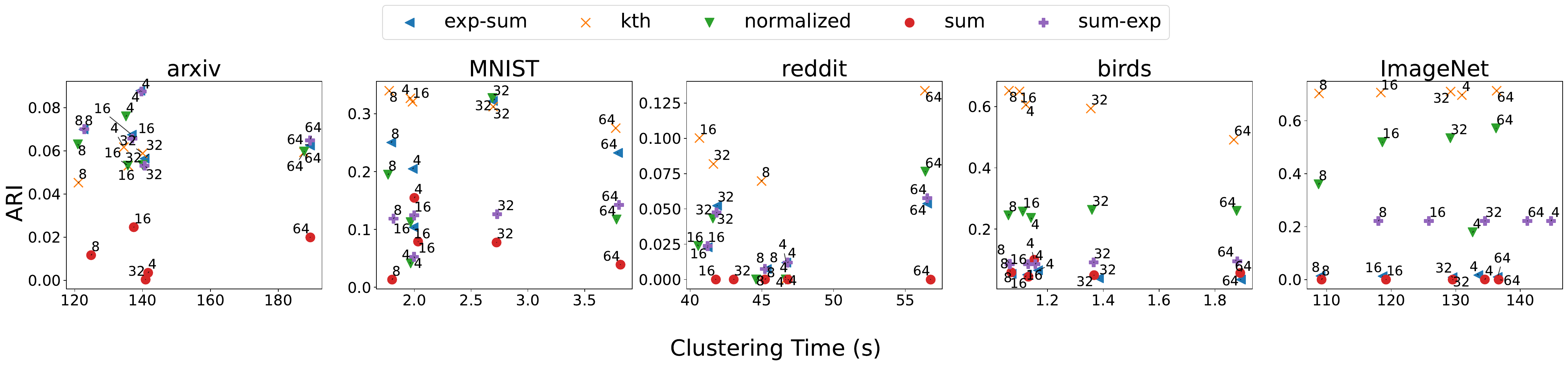}
  \caption{Clustering quality (ARI) vs.\ runtime of \framework when using different density functions and values of $k$. The $y$-axis shows the ARI scores computed with respect to the ground truth. The $x$-axis shows the runtime in seconds. Each color represents a density function, and the number next to each data point is the value of $k$ used.} 
  \label{fig:ari_vs_time}
\end{figure*}

\subsection{Comparison of Different Density Functions, Values of $k$, and Graph Indices}
In \Cref{fig:ari_vs_time}, we show the runtime vs.\ ARI of different density functions and values of $k$. We see that the \texttt{kth} density function is the most robust and achieves the highest ARI score on most datasets. We also observe that using $k=16$ provides a good trade-off between quality and time. 
\texttt{exp-sum}, \texttt{sum}, and \texttt{sum-exp} are other density functions in \framework, which are combinations of the distances to the $k$-nearest neighbors. We describe them in our full paper.

\begin{figure}[t]
 \centering
 \hfill
 \includegraphics[width=0.45\columnwidth]{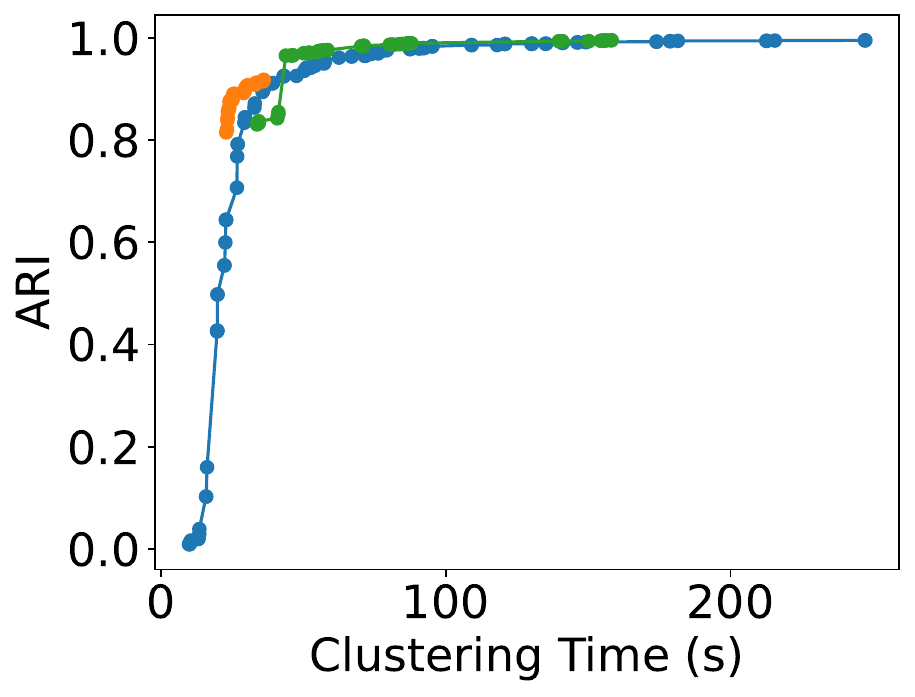}
 \hfill
 \includegraphics[width=0.45\columnwidth]{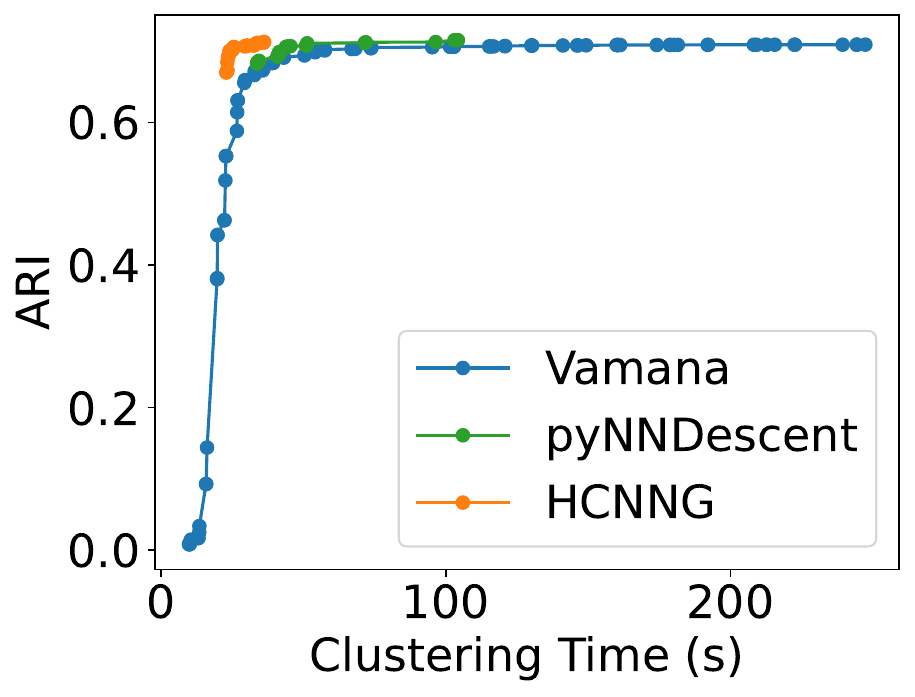}
 \hfill
  \caption{\textbf{(Left)} Pareto frontier of clustering quality with respect to \textit{exact DPC} vs.\ runtime on \datasetname{ImageNet}. \textbf{(Right)} Pareto frontier of clustering quality with respect to the \textit{ground truth clustering} vs.\ runtime on \datasetname{ImageNet}.}
  \label{fig:different_graphs}
\end{figure}

We can easily swap in different graph indices into our framework and compare the results. In \Cref{fig:different_graphs}, we show a Pareto frontier of the clustering quality vs.\ runtime on \datasetname{ImageNet} for each of the following different graph indices: \algname{Vamana}~\cite{jayaram2019diskann}, \algname{pyNNDescent}~\cite{pynndescent}, and \algname{HCNNG}~\cite{munoz2019hcnng}. The Pareto frontier comprises points that are non-dominated, meaning no point on the frontier can be improved in quality without worsening time and vice versa. In other words, the curve we plot represents the optimal trade-off in the parameter space between clustering time and quality. 

To create the Pareto frontier, we do a grid search for each method over different choices of maximum degree $R$ and the beam sizes for construction, \knn search, and dependent point finding. We choose all combinations of these four parameters from $[8, 16, 32, 64, 128, 256]^4$. We set the density method to be \texttt{kth} without normalization and $k = 16$.
We set $\alpha = 1.1$ for \algname{Vamana} and \algname{PyNNDescent}. \algname{HCNNG} and \algname{PyNNDescent} additionally accept a $num\_repeats$ argument, which represents how many times we independently repeat the construction process before merging the results together; we set this parameter equal to $3$.
We see that all graph indices are able to achieve similar maximum ARI with respect to the ground truth: \algname{Vamana}, \algname{HCNNG}, and \algname{PyNNDescent} achieve maximum ARIs of $0.709$, $0.715$, and $0.713$, respectively. \algname{HCNNG} attains this maximum slightly faster than the other two indices, but when compared to the exact DPC result, \texttt{HCNNG} has a smaller maximum ARI, which means its clustering deviates more from the exact solution. Indeed, \algname{HCNNG} has a maximum ARI compared to exact DPC of $0.918$, while \algname{PyNNDescent} and \algname{Vamana} attain a maximum ARI of $0.995$ compared to exact DPC. 

We also find that among the four Vamana hyperparameters, the maximum degree of the graph and construction beam size have both the largest contribution to the ARI and the largest impact on the clustering time.
\iffull
Please find more details in \Cref{sec:appendix-exp}.
\fi

\begin{figure*}[t]
\vspace{-5pt}
  \centering \includegraphics[width=0.95\textwidth]{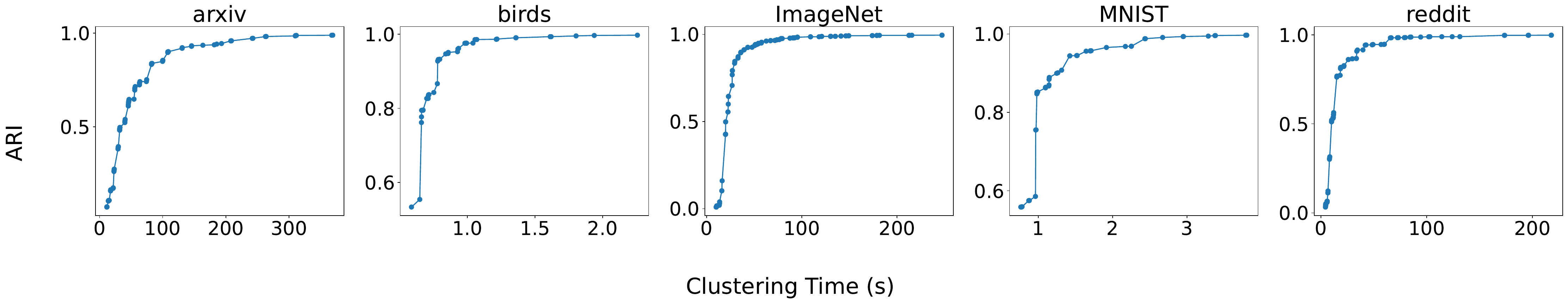}
  \vspace{-5pt}
  \caption{Pareto frontier of clustering quality of \framework with respect to exact DPC.}
  \label{fig:pareto_bruteforce}
\end{figure*}

\begin{figure*}[t]
  \centering \includegraphics[width=0.95\textwidth]{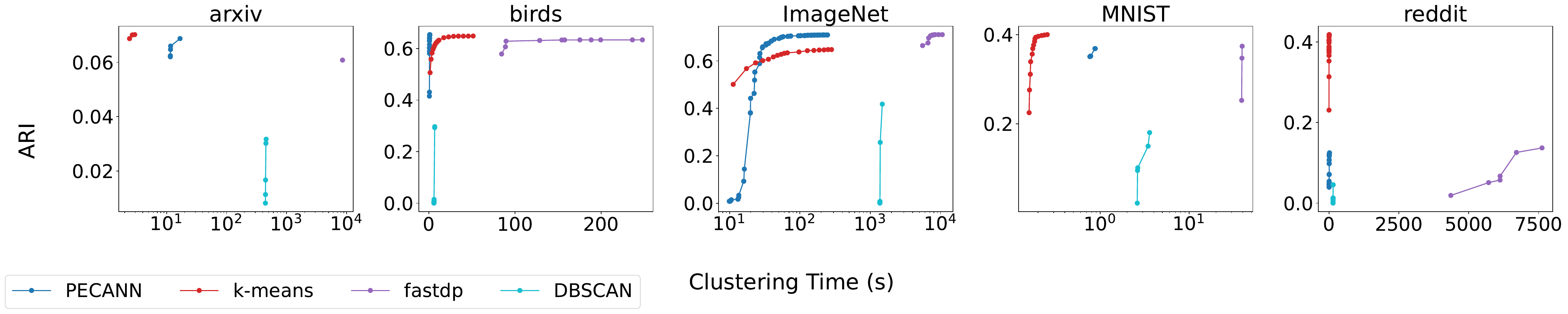}
    \vspace{-3pt}
  \caption{Pareto frontier of ARI with respect to ground truth vs.\ runtime. Up and to the left is better. 
  \framework is the best method on \datasetname{ImageNet} and \datasetname{birds}, has similar performance to the best method (\algname{$k$-means}) on \datasetname{arxiv}, and is slower or has worse quality than the best method (\algname{$k$}-means) on \datasetname{mnist} and \datasetname{reddit}.
  \algname{fastdp} is sequential.
  The $x$-axis on \datasetname{arxiv}, \datasetname{ImageNet}, and \datasetname{MNIST} are in log-scale.
  }
  \label{fig:pareto_groundtruth}
\end{figure*}

\subsection{Clustering Quality-Time Trade-off}
In \Cref{fig:pareto_bruteforce}, we plot the Pareto frontier of clustering quality (ARI with respect to the exact DPC clustering) vs.\ runtime of \framework. To obtain the Pareto frontiers, we use the same parameter values as in the last experiment, except that for the smaller datasets with $n < 250,000$ we use a smaller range $[8, 16, 32, 64]^4$ for the parameter search space.
We see that \framework can achieve results very close to the exact DPC clustering. On all datasets except \datasetname{arxiv}, \framework achieves at least $0.995$ ARI with respect to exact DPC, and on \datasetname{arxiv}, \framework achieves $0.989$ ARI with respect to exact DPC.

\subsection{Comparison of Different Methods}\label{sec:compare-methods}

In \Cref{fig:pareto_groundtruth}, we plot the Pareto frontier of clustering quality (ARI with respect to the ground truth clustering) vs.\ runtime for different methods on the larger datasets. 
To obtain the Pareto frontiers, we use the same parameters for \algname{Vamana} as in the previous experiment. 
For \algname{k-means}, we use $nredo \in [1, 2, 3, 4]$ and $niter \in [1, 2, 3, \ldots, 9, 10, 15, 20, 25, \ldots, 40, 45]$, for all combinations where $niter \times nredo < 100$. For \algname{fastdp}, we use $window\_size \in [20, 40, 80, 160, 320]$ for all datasets (controlling query quality) and $max\_iterations \in [1, 2, 4, 8, 16, 32, 64]$ (controlling graph construction quality). 
For \algname{DBSCAN}, we use different parameters for each dataset, based on guidelines from~\cite{sander1998density,schubert2017dbscan,rahmah2016determination}. \cite{sander1998density} suggest setting $min\_pts$ to $2d-1$. For high-dimensional datasets, \cite{schubert2017dbscan} suggest that increasing $min\_pts$ may improve results. $\epsilon$ is chosen based on the distribution of the $min\_pts$-nearest neighbor distances~\cite{rahmah2016determination}. The parameters
can be found in 
\iffull
\Cref{sec:parameters}.
\fi

We observe that \algname{DBSCAN} has lower quality and higher runtime than all other baselines. As the original authors of DBSCAN state, it is difficult to use DBSCAN for high-dimensional data~\cite{schubert2017dbscan}.

We observe that the sequential \algname{fastdp} is slower that \framework on all datasets. 
In terms of accuracy, \framework has better maximum ARI on \datasetname{birds} and \datasetname{arxiv}, while \algname{fastdp} has better maximum ARI on \datasetname{reddit} (although as we discuss below, \datasetname{reddit} is not well suited to DPC).

Compared with \algname{$k$-means}, \framework obtains better quality and is faster on \datasetname{ImageNet} and \datasetname{birds}, where the number of ground truth clusters is large, and performs about equal with \algname{$k$-means} on \datasetname{arxiv}. However, \framework has worse quality for a given time limit on \datasetname{reddit} and \datasetname{mnist}. 
Although \framework has worse quality than \algname{$k$-means} on two datasets when \algname{$k$-means} uses the correct number of clusters, \algname{$k$-means}'s quality is sensitive to the number clusters. As shown in \Cref{sec:varying-num-clusters}, \algname{$k$-means} can have lower quality than \framework on these two datasets when $k$ 
is not the number of ground truth clusters. 

We summarize the best ground truth ARI and the corresponding parallel running time that all these methods, as well as \algname{BruteForce}, achieve in \Cref{tab:different_graph_methods}. Compared to density-based methods, \framework achieves 37.7--854.3x speedup over \algname{BruteForce}, 45--734x speedup over \algname{fastdp},
while achieving comparable ARI. \framework also achieves up to 0.7 higher ARI than \algname{DBSCAN}, and is up to orders-of-magnitude faster.

For more intuition on the runtime differences between \framework and \algname{$k$-means}, note that the work of each iteration of \algname{$k$-means} is linear in the number of clusters multiplied by $n$, and so \algname{$k$-means} is fast on datasets like \datasetname{MNIST} with a small number of ground truth clusters, while it is slower on datasets like \datasetname{birds} and \datasetname{ImageNet} that have many clusters. 

In terms of an explanation for the quality difference between \framework and \algname{$k$-means}, \framework gets better maximum accuracy on \datasetname{ImageNet} and \datasetname{birds}, which may be because the ground truth clusters in these datasets form shapes that our density-based method \framework can find, but that the geometrically constrained $k$-means cannot. On the other hand, for \datasetname{reddit},
\framework has lower quality than \algname{$k$-means}. Since we still obtain cluster quality very close to the exact DPC on \datasetname{reddit} (see \Cref{fig:pareto_bruteforce}), this dataset is a case where the density based DPC method is worse than the simpler $k$-means heuristic.

\begin{table}[!t]
\footnotesize
\centering
 \begin{tabular}{l c c c} 
 \toprule
 Algorithm & Dataset & Time (s) & Maximum ARI  \\ 
\midrule
\framework & \datasetname{arxiv} & 11.65 & 0.07\\
\algname{fastdp} & \datasetname{arxiv} & 8557.89 & 0.06\\
\algname{BruteForce} & \datasetname{arxiv} & 9953.15 & 0.07\\
\algname{kmeans} & \datasetname{arxiv} & 2.41 & 0.07\\
\algname{DBSCAN} & \datasetname{arxiv} & 451.99 & 0.03\\
\midrule
\framework & \datasetname{birds} & 0.86 & 0.65\\
\algname{fastdp} & \datasetname{birds} & 128.71 & 0.63\\
\algname{BruteForce} & \datasetname{birds} & 66.04 & 0.66\\
\algname{kmeans} & \datasetname{birds} & 28.66 & 0.65\\
\algname{DBSCAN} & \datasetname{birds} & 6.79 & 0.30\\
\midrule
\framework & \datasetname{ImageNet} & 101.58 & 0.71\\
\algname{fastdp} & \datasetname{ImageNet} & 7655.91 & 0.71\\
\algname{BruteForce} & \datasetname{ImageNet} & 31979.98 & 0.71\\
\algname{kmeans} & \datasetname{ImageNet} & 188.17 & 0.65\\
\algname{DBSCAN} & \datasetname{ImageNet} & 1481.39 & 0.42\\
\midrule
\framework & \datasetname{MNIST} & 0.87 & 0.37\\
\algname{fastdp} & \datasetname{MNIST} & 39.36 & 0.37\\
\algname{BruteForce} & \datasetname{MNIST} & 32.80 & 0.34\\
\algname{kmeans} & \datasetname{MNIST} & 0.22 & 0.40\\
\algname{DBSCAN} & \datasetname{MNIST} & 3.59 & 0.18\\
\midrule
\framework & \datasetname{reddit} & 14.90 & 0.12\\
\algname{fastdp} & \datasetname{reddit} & 7621.71 & 0.14\\
\algname{BruteForce} & \datasetname{reddit} & 2888.20 & 0.10\\
\algname{kmeans} & \datasetname{reddit} & 5.36 & 0.42\\
\algname{DBSCAN} & \datasetname{reddit} & 148.48 & 0.05\\
 \bottomrule
\end{tabular}
\caption{\label{tab:different_graph_methods} The maximum ARI score with respect to the ground truth achieved by different clustering algorithms across different datasets, and their corresponding parallel running time. 
}
\label{tab:max_ari_scores}
\end{table}

\subsection{Varying Number of Clusters} \label{sec:varying-num-clusters}

\begin{figure}[t]
 \centering
 \includegraphics[width=0.8\columnwidth]{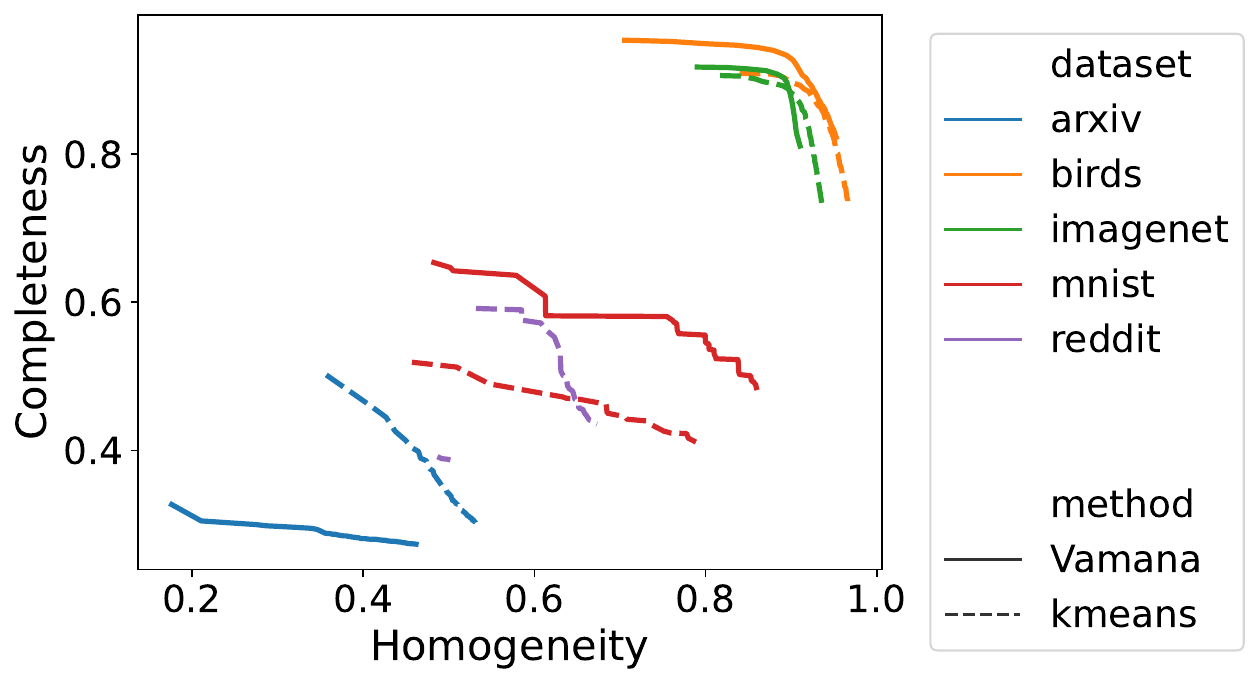}
   \vspace{-5pt}
  \caption{Pareto frontiers of completeness vs.\ homogeneity of \framework and \algname{$k$-means} on different datasets. Up and to the right is better. }
  \label{fig:compelteness}
\end{figure}

Not knowing the number of ground truth clusters is common in real-world settings. %
In \Cref{fig:compelteness}, we show a Pareto frontier of the completeness and homogeneity scores (with respect to ground truth) of \framework and \algname{$k$-means} on different datasets with varying number of clusters. We generate this Pareto frontier using the same experiment setup as earlier, except  now we record homogeneity and completeness instead of ARI as we vary the number of clusters given to each method. Thus, points along the Pareto frontier in \Cref{fig:compelteness} are optimal tradeoffs between homogeneity and completeness as we vary the cluster granularity. We see that \framework strictly dominates \algname{$k$-means} on \datasetname{birds} and \datasetname{MNIST}, and \algname{$k$-means} is better on \datasetname{arxiv} and \datasetname{reddit}. On \datasetname{ImageNet}, \framework achieves higher completeness and \algname{$k$-means} achieves higher homogeneity.

\iffull
We also study the ARI of \framework using \algname{Vamana} and \algname{$k$-means} when we pass a number of clusters to the algorithm different than the ground truth in
\Cref{sec:appendix-varying-num-clusters}.
\fi
When the number of clusters used is larger than the ground truth, the quality of \algname{$k$-means} decays quickly while the quality of \framework is more robust.

%% file: related_work.tex
\section{Related Work}\label{sec:related}
\iffull
In this section, we give an overview of the different DPC variations that have been proposed since the original DPC algorithm~\cite{Rodriguez14}, particularly the ones that are based on \knn{s}. We also briefly introduce other density-based clustering algorithms. Finally, we discuss recent advances on graph-based ANNS. 
\fi

\myparagraph{Variants of DPC}
The original DPC algorithm~\cite{Rodriguez14} uses a range search to compute the density of a point $x$, where the density is defined as the number of points in a ball of fixed radius centered at $x$.
In contrast, while \framework supports any density metric, our paper focuses specifically on \knn-based DPC variants, which do not require a range search. These methods are less sensitive to noise and outliers~\cite{sddp} and are more computationally efficient to compute in high dimensions. 
Some of these methods (e.g.,~\cite{sddp, yaohui2017adaptive, sun2021density, xie2016robust}) also have a refinement step after obtaining the initial DPC clustering. For these methods, \framework can be used to efficiently obtain the first DPC clustering before the refinement step.

Floros et al.~\cite{sddp} and Chen et al.~\cite{chen2020fast} use the inverse of the distance to the $k^\text{th}$ nearest neighbor as the density measure. 
Sieranoja and Fr{\"a}nti~\cite{sieranoja2019fast} propose \algname{FastDP}, which uses the inverse of the average distance to all \knn{s} as the density measure, and finds the \knn{s} by constructing an approximate \knn graph.
\iffull
Their motivation for not using the original DPC density function is that they are considering non-metric distance measures. Specifically, they use string similarity measures such as the Levenshtein distance and the Dice coefficient. In our experiments, we used the Euclidean distance measure for \algname{FastDP} to be consistent. 
\fi
d’Errico et al.~\cite{d2021automatic} propose a variant of DPC for high-dimensional data. It combines DPC with a non-parametric density estimator called PAk, but their algorithm is sequential.
\iffull
Du et al.~\cite{du2018density} propose a density function that depends on the shortest path distance in the \knn graph. 
Some works propose to normalize the density of each point by the density of its neighbors~\cite{su2018bpec, hou2019enhancing, geng2018recome}, as the normalization helps to reduce the influence of large density differences across clusters and is better for detecting clusters with different densities~\cite{hou2019enhancing}. 
Yin et al.~\cite{yin2022improved} use \knn searches to partition the data, which they show works well when the average density between clusters is very different.
There are also works that propose to use density measures based on the mutual neighborhood~\cite{sun2021density, li2018comparative, chen2022improved}, natural neighborhood~\cite{ding2023improved, zhu2016natural}, order similarity~\cite{yang2021gdpc}, and the exponential of the sum of distances to neighbors~\cite{li2018comparative, xie2016robust, du2016study, yaohui2017adaptive}.

\fi
There are also algorithms that perform dimensionality reduction on the dataset before running DPC~\cite{du2016study, chen2022improved}.

\myparagraph{Parallel, Approximate, and Dynamic DPC}
Zhang et al.~\cite{Zhang16} propose an approximate DPC algorithm for MapReduce using locality-sensitive hashing.
\iffull
It partitions the data set into buckets, and searches within relevant buckets to find approximate dependent points. It resorts to scanning the whole dataset when the approximate dependent point does not seem accurate.
\fi
Amagata and Hara~\cite{Amagata21} propose a partially parallel exact DPC algorithm and two parallel grid-based approximate DPC algorithms. 
\iffull
They show that their algorithms are faster than previous solutions, including LSH-DDP~\cite{Zhang16}, CFSFDP-A~\cite{bai2017fast}, FastDPeak~\cite{chen2020fast}, and DPCG~\cite{xu2018dpcg}.
\fi
They also propose parallel static and dynamic
DPC algorithms for data in Euclidean space~\cite{amagata2023efficient, amagata2022scalable}. 
\iffull
Amagata et al.\cite{amagata2023efficient} show that their dynamic algorithm outperforms previous dynamic DPC algorithms~\cite{ulanova2016clustering, gong2017clustering}.
\fi
Huang et al.~\cite{dpc} propose a parallel exact DPC algorithm based on priority \kdt{s} and show their algorithm outperforms previous tree-index approaches~\cite{Amagata21, Rasool20}. 
Lu et al.~\cite{lu2020distributed} propose speeding up DPC using space-filling curves.
Unlike \framework, these algorithms~\cite{Amagata21, amagata2023efficient, dpc, bai2017fast} are only efficient on low-dimensional datasets and must be used with Euclidean distance.

Amagata~\cite{amagata2022scalable} proposes an approximate dynamic DPC algorithm for metric data, but it is sequential and only tested on datasets with up to 115 dimensions. In comparison, \framework is parallel and we experimented on datasets with up to 1024 dimensions. 
There are also dynamic algorithms for \knn-based DPC variants~\cite{seyedi2019dynamic, du2022dynamic}.

\myparagraph{Density-based Clustering Algorithms}
DPC falls under the broad category of density-based clustering algorithms, which have the advantage of being able to detect clusters of arbitrary shapes. 
Some density-based clustering algorithms define the density of a point based on the number of points in its vicinity~\cite{Ester96,Agrawal98,Ankerst99,Januzaj04,Rodriguez14, sddp, Chen19}. Others use a grid-based definition, which first quantizes the space into cells and then does clustering on the cells~\cite{Wang97,Hinneburg98,Hanmanthu18,Sheikholeslami00}. 
Still others use a probabilistic density function~\cite{Wang97,Kriegel05,Smiti13}. One popular density-based clustering algorithm is DBSCAN~\cite{Ester96}, which has many derivatives as well~\cite{Ankerst99,Tepwankul10,Gotz15,Borah04,Ertoz03,Campello2015, chen2020knn}. 
However, the original authors of DBSCAN state that it is difficult to use for high-dimensional data~\cite{schubert2017dbscan}. 

\myparagraph{Graph-based Approximate Nearest Neighbor Search (ANNS)}
Graph-based ANNS methods have been shown to be effective in practice~\cite{wang2021comprehensive, ANNScaling, wanggraph}. Existing graph-based indices include Hierarchical Navigable Small World Graph (HNSW)~\cite{Malkov2020ann}, DiskANN (also called Vamana)~\cite{jayaram2019diskann}, HCNNG~\cite{munoz2019hcnng}, PyNNDescent~\cite{pynndescent}, $\tau$-MNG~\cite{peng2023efficient}, and many others (e.g.,~\cite{zhao2023towards, chen2023finger, coleman2022graph}). Please see \cite{ANNScaling} and \cite{wang2021comprehensive} for comprehensive overviews of these methods and their comparisons with non-graph-based methods, such as locality-sensitive hashing, inverted indices, and tree-based indices. 

The dependent point search in DPC can also be viewed as a filtered search, where the points' labels are their density, and we filter for points with densities larger than the query point's density. Various graph-based similarity search algorithms have been adapted recently to support filtering~\cite{zhao2022constrained, wang2022navigable, gollapudi2023filtered, gupta2023caps}.
Gollapudi et al.~\cite{gollapudi2023filtered} propose the Filtered DiskANN algorithm, which supports filtered ANNS queries, where nearest neighbors returned must match the query's labels.
Gupta et al.~\cite{gupta2023caps} developed the CAPS index for filtered ANNS via space partitions, which supports conjunctive constraints while DiskANN does not. Both DiskANN~\cite{singh2021freshdiskann} and CAPS can be made dynamic.
However, these solutions use categorical labels, and a point can have multiple labels. 
Using this approach for dependent point finding requires quadratic memory just to specify the labels (the $i^\text{th}$ least dense point would need $i-1$ labels, which are the $i-1$ smaller density values than its density), which is prohibitive. Indeed,
we tried running the Filtered DiskANN code on our datasets but it ran out of space on our machine.
VBASE~\cite{Zhang2023} also supports filtered search by first searching for  $k$-nearest neighbors and then filtering. However, they do not 
handle the case when there are no neighbors returned that satisfy the criteria.

\iffull
There are also works that explore the theoretical aspects of graph-based ANNS~\cite{navarro2002searching, prokhorenkova2020graph, shrivastava2023theoretical, laarhoven2018graph, indyk2023worstcase}. It is known that to find the exact nearest neighbor for any possible query via a greedy search, the graph must contain the Delaunay graph as a subgraph. Unfortunately, Delaunay graphs have high degrees in high dimensions and cannot be constructed efficiently~\cite{navarro2002searching, prokhorenkova2020graph}. 
Laarhoven~\cite{laarhoven2018graph} provides bounds for nearest neighbor search on datasets uniformly distributed on a $d$-dimensional sphere with $d \gg \log n$ and provides time–space trade-offs for ANNS.  
Prokhorenkova and Shekhovtsov~\cite{prokhorenkova2020graph} extend this work and analyze the performance of graph-based ANNS algorithms in the low-dimensional ($d \ll \log n$) regime. 
Peng et al.~\cite{peng2023efficient} propose a new graph index and prove that if the distance between a query and its nearest neighbor is less than a constant, the search on their graph is guaranteed to find the exact nearest neighbor and the time complexity of the search is small. Indyk and Xu~\cite{indyk2023worstcase} study the worst-case performance of graph-based ANNS algorithms, including DiskANN, HNSW, and NSG. 
They show non-trivial bounds on accuracy and query time for a "slow preprocessing" version of DiskANN, and provide examples of poor worst-case behavior for the regular version of DiskANN, HNSW, and NSG.

There has also been work that uses approximate nearest neighbor oracles for other clustering problems~\cite{ullah2022clustering}. 
\fi

%% file: conclusion.tex
\section{Conclusion}
We present the \framework framework for density peaks clustering (DPC) variants in high dimensions. We adapt graph-based approximate nearest neighbor search methods to support (filtered) proximity searches in DPC variants.   
\framework is highly parallel and scales to large datasets. 
We show several DPC variants that can be implemented in \framework, and evaluate them on large datasets. \framework achieves significant improvements in runtime and clustering quality over the state of the art.

%% file: appendix-functions.tex
\section{Additional Functions in \framework}\label{sec:appendix-functions}

\subsection{Density Functions}
\framework provides the following density functions in addition to the ones presented in \Cref{sec:density}.

\myparagraph{\texttt{exp-sum}}
The density of $x_i$ is  $\rho_i = \exp(-\frac{\sum_{j \in \mathcal{N}_i} D(x_i, x_j)^2}{k})$, which is exponential in the negative of the average squared distance between $x_i$ and its $k$-nearest neighbors~\cite{du2016study}. Each density computation takes $O(k)$ work and $O(\log k)$ span by using a parallel sum.

\myparagraph{\texttt{sum-exp}} The density of $x_i$ is 
 $\rho_i = \frac{\sum_{j \in \mathcal{N}_i} \exp(-D(x_i, x_j)^2)}{k}$~\cite{yaohui2017adaptive}. Each density computation is $O(k)$ work and $O(\log k)$ span to compute the summation using a parallel sum. It can also be viewed as a variant of the kernel density of the original DPC algorithm~\cite{Rodriguez14}.

\myparagraph{\texttt{sum}} The density of $x_i$ is 
 $\rho_i = -\sum_{j \in \mathcal{N}_i} D(x_i, x_j)$, which is the negative sum of distances to the $k$-nearest neighbors. Each density computation takes $O(k)$ work and $O(\log k)$ span to compute the summation using a parallel sum. We include this as a simple baseline.

\subsection{Center Functions }
\label{sec:appendix-center}
We describe the additional center functions $F_{\text{center}}$ that we implement in \framework. Recall from \Cref{sec:prelim} that $\delta_i =  D(x_i,\lambda_i)$ is the dependent distance of $x_i$.

\myparagraph{Local Center} For this method, a point is a center point if it has the highest density among its \knn{s}. 
This can be implemented using a parallel filter with $O(nk)$ work and $O(\log n)$ span. This density finder is usually accompanied by further steps to merge and refine the initial DPC clusters~\cite{sun2021density, sddp}.

There are also methods that plot a decision graph for visualization and pick the cluster centers manually~\cite{du2018density, du2018robust}.
Finally, some iterative methods have been proposed (e.g.,~\cite{xie2016robust, abbas2021denmune, liu2018shared}), but they have been infrequently used.

\subsection{Noise Function}
This noise function described in \Cref{sec:density} is the only one that \framework currently implements because it is the most commonly used, but alternative noise function definitions can be supported.
For example, \cite{abbas2021denmune} define noise points based on
the number of neighborhoods a point belongs in, and
\cite{xie2016robust} compute noise points by using a threshold on the dependent distance.

%% file: appendix-analysis.tex
\section{Approximation Analysis}\label{sec:appendix-approximation-analysis}

\subsection{Density Estimation}

We analyze the density approximated by the \texttt{kth} density function assuming that we can find $c$-approximate $k$-nearest neighbors. 

\begin{definition}[$c$-approximate $k$-nearest neighbors]~\label{def:knn}
Let $p_j$ be the true $j^\text{th}$ nearest neighbor of query point $q$.
Let $\mathcal{N}$ be the returned set of approximate $k$-nearest neighbors of $q$. 
Let $\hat{p}_j$ be the point in $\mathcal{N}$ that is $j^\text{th}$ furthest from $q$,

$\mathcal{N}$ is $c$-approximate $c \geq 1$ if (1) for all $j\leq k$, $D(q, p_j) \leq D(q, \hat{p}_j) \leq c \cdot D(q, p_j)$,
and (2) $\{p' : D(p', q) \leq \frac{D(p_k, q)}{c} \} \subseteq \mathcal{N}$, i.e., the set of points within distance $\frac{D(p_k, q)}{c}$ to $q$ is a subset of $\mathcal{N}$. 
\end{definition}
The first condition guarantees that the furthest point in the approximate $k$-nearest neighbors are not too far from the true $k$-nearest neighbors.
Note that for some density functions, the first condition can be weaker. For example, the \texttt{kth} density function only requires this condition when $j=k$.
The second condition guarantees that the 
points that are sufficiently close
to the query point are returned among the approximate $k$-nearest neighbors.

\begin{definition}[Density Interval]
    The \textit{density interval} of a point $q$ is a range that gives the lower and upper bounds of the approximate density of $q$. 
\end{definition}

Let $r_q$ be the distance between $q$ and its $k$-nearest neighbor.
If we use an algorithm that guarantees $c$-approximate $k$-nearest neighbors, point $q$ has density interval $[\frac{1}{cr_q}, \frac{1}{r_q}]$. Consider all points $[1, \dots, n]$, ordered from having the highest true density to having the lowest true density. Consider the list of intervals $[[\frac{1}{cr_1}, \frac{1}{r_1}], \dots, \allowbreak [\frac{1}{cr_n}, \frac{1}{r_n}]]$ in the same order (note that we are only using this order for analysis, and our algorithm does not need to compute this order).
\begin{definition}[Conflict]
A point $q_i$'s density range $[a_i, b_i]$ has a \textit{conflict} with another point $q_j$'s density range $[a_j, b_j]$ if $[a_i, b_i]$ and $[a_j, b_j]$ has any overlap. For $i < j$, conflict happens when $a_i < b_j$.
\end{definition}

If the list of intervals does not conflict, our density estimation does not affect the correctness of subsequent steps, as only the relative ranking of densities is used when identifying dependent points. Moreover, if a contiguous chunk of points $[x_i, \dots, x_j]$ have conflicts, these overlaps only affect the dependent point search for the points $[x_i, \dots, x_j]$, and not points before $i$ and after $j$ in the ordering.

The following lemma guarantees that the density peaks of the exact algorithm that do not conflict with other points will remain density peaks. 

\densitylemma*
\begin{proof}
   A point $q$ is a density peak with threshold $\delta_{\min}$ if $q$'s distance to its dependent point is greater than $\delta_{\min}$.
   Since there is no conflict with $q$'s interval, the set of points with higher density than $q$ is the same as in the exact algorithm. As a result, $q$ can only find an approximate nearest neighbor that is either the same distance from or further away from its exact dependent point. Therefore, its distance to the approximate dependent point must be at least as large by Definition~\ref{def:knn} and it stays a density peak. 
\end{proof}

Note that there may be additional density peaks returned by the approximate algorithm, but the true density peaks in the exact algorithm are guaranteed to still be density peaks.

\subsection{Dependent Point Estimation}

Now we analyze the approximate dependent point found by \Cref{alg:computedeppts}. 
The following lemma guarantees that the approximate dependent points returned by our algorithm are not too much further than the true dependent points.
Let $d_j$ be the distance to the true $j^\text{th}$ nearest neighbor from query point $q$.

\dependentpointlemma*
\begin{proof}
When we find an approximate $k$-nearest neighbor, everything within $\frac{d_k}{c}$ has been found by \Cref{def:knn}, so if we have not found a dependent point of query point $q$, it must be at least $\frac{d_k}{c}$-away from $q$.
Suppose we found our approximate parent within the approximate $\beta k$-nearest neighbor which has a distance at most $cd_{\beta k}$ to $p$. Then, the approximate dependent point is at most $c^2 \frac{d_{\beta k}}{d_k}$ times further from $q$ than the exact dependent point. 
\end{proof}

In \Cref{alg:computedeppts}, we use $\beta=2$
 for \Cref{lemma:dependent-point}, since we double the number of nearest neighbors to find until we have found a dependent point.

%% file: appendix-exp.tex
\section{Additional Experiments}\label{sec:appendix-exp}
\datasetname{S2} and \datasetname{Unbalanced}~\cite{franti2018k} are small 2-dimensional baseline datasets used in prior clustering papers.
We summarize the datasets in \Cref{tab:datasets-appendix} and describe the parameters we used for them in \Cref{tab:params-appendix}.
For DBSCAN, we used Wang et al.'s~\cite{wang2019dbscan} parallel C++ implementation, which is optimized for low-dimensional data sets, instead of scikit-learn. The other algorithms are the same as described in \Cref{sec:exp-setup}.

\begin{table}[t]
\footnotesize
\centering
 \begin{tabular}{l c c c c c} 
 \toprule
 Name & $n$ & $d$ & Description & \# Clusters\\ %
 \midrule
 \datasetname{S2} & 5,936 & 2 & Standard benchmark  & 15\\ 
 \datasetname{Unbalanced} & 6,500 & 2 & Standard benchmark & 8\\ 
 \bottomrule
\end{tabular}
\caption{\label{tab:datasets-appendix}Small datasets used in our experiments.}
\end{table}

\begin{table}[!t]
\footnotesize
\centering
\begin{tabular}{c c c c c }
\toprule
Dataset & $L$ & $L_d$ & $R$ & $k$\\ \hline
\datasetname{S2, Unbalanced} & 12 & 4 & 16 & 6\\
\end{tabular}
\caption{Default parameters used for the small datasets.
}
\label{tab:params-appendix}
\end{table}

We show in Table~\ref{tab:small_experiment} the runtime and ARI score with respect to the ground truth of all methods run using a single thread on the small datasets \datasetname{S2} and \datasetname{Unbalanced}. Compared to the density-based methods, \algname{$k$-means} has a slightly higher ARI on \datasetname{S2}, but significantly worse ARI on \datasetname{Unbalanced}. This shows that the relative performance of \algname{$k$-means} and density-based methods depends on the dataset, which we also observed on large high-dimensional real-world datasets. 

\begin{table}[!t]
\footnotesize
\centering
\setlength{\tabcolsep}{3pt}
\begin{tabular}{l c c c c c c} 
\toprule
Dataset & $k$ & Index & Density & Dependent Point & Union-Find \\
\midrule
\datasetname{arxiv} & 8 & \textbf{60.9} & 35.0 & 4.1 & 0.0\\
\datasetname{arxiv} & 32 & \textbf{52.5} & 45.6 & 1.9 & 0.0\\
\midrule
\datasetname{MNIST} & 8 & \textbf{53.4} & 43.7 & 2.6 & 0.3\\
\datasetname{MNIST} & 32 & 34.8 & \textbf{63.7} & 1.4 & 0.1\\
\midrule
\datasetname{reddit} & 8 & \textbf{54.6 }& 36.9 & 8.4 & 0.1\\
\datasetname{reddit} & 32 & \textbf{58.6 }& 40.1 & 1.2 & 0.1\\
\midrule
\datasetname{birds} & 8 & \textbf{72.0} & 20.8 & 6.0 & 1.2\\
\datasetname{birds} & 32 & \textbf{55.2 }& 38.8 & 5.1 & 0.9\\
\midrule
\datasetname{ImageNet} & 8 & \textbf{70.0} & 27.3 & 2.6 & 0.1\\
\datasetname{ImageNet} & 32 & \textbf{59.0} & 39.0 & 2.0 & 0.0\\
\bottomrule
\end{tabular}
\caption{\label{tab:small_runtime} Runtime percentage breakdown with the \texttt{kth} density method with $k=8,32$ on large datasets. 
}
\end{table}

\subsection{Hyperparameter Regression Analysis}
We also run a linear regression for each of our five main datasets to predict the clustering time and the ARI from the four Vamana hyperparameters: the maximum degree $R$ for graph construction, and the three beam size hyperparameters for graph construction, \knn search, and dependent point finding. We use the log of each of the hyperparameters for the ARI regression. Averaging across the five regressions, the ARI regressions have an average $R^2$ of $0.714$ and the hyperparameters have average linear regression weights $0.125$, $0.139$, $7.69\mathrm{e}{-3}$, and $1.25\mathrm{e}{-3}$, respectively, while the clustering time regressions have an average $R^2$ of $0.783$ and average weights of $0.181$, $0.360$, $0.0429$, and $1.37\mathrm{e}{-4}$, respectively.  In summary, the maximum degree of the graph and construction beam size have both the largest contribution to the ARI and the largest impact on the clustering time.

\begin{figure*}[t]
\vspace{-5pt}
  \centering \includegraphics[width=0.95\textwidth]{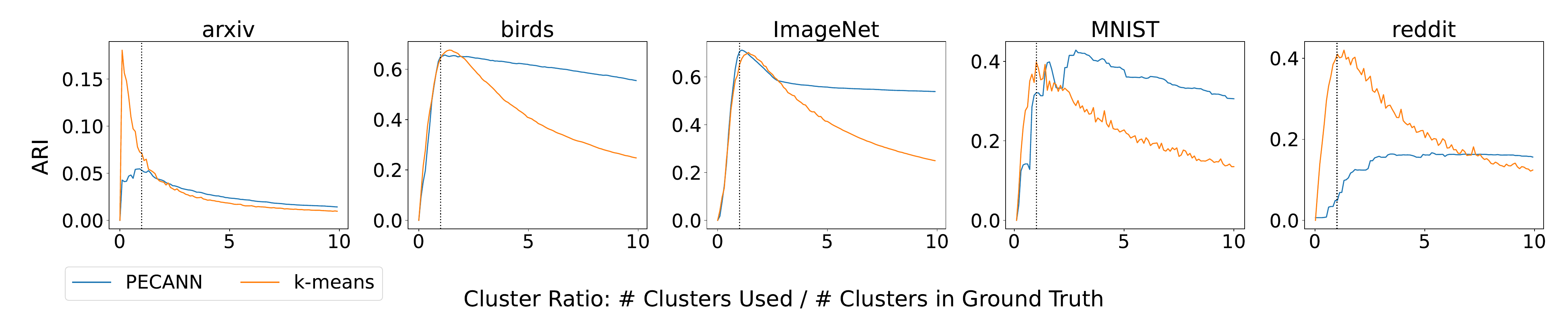}
  \vspace{-3pt}
  \caption{The ARI of \framework using the \algname{Vamana} graph index vs.\ that of \algname{$k$-means}, when clustering with different numbers of clusters than the ground truth. $y$-axis is the ARI with respect to the ground truth. $x$-axis is the ratio between the number of clusters used and the number of clusters in the ground truth. The vertical dotted line is $x=1$, where the correct number of clusters is used. }
  \label{fig:varying_num_cluster}
\end{figure*}

\subsection{Varying Number of Clusters} \label{sec:appendix-varying-num-clusters}
In \Cref{fig:varying_num_cluster}, we show the ARI (with respect to ground truth) of \framework using \algname{Vamana} and \algname{$k$-means} when we pass a number of clusters to the algorithm different than the ground truth (we plot the ratio between the number of clusters used and the number of ground truth clusters). Not knowing the number of ground truth clusters is common in real-world settings, so algorithm performance in this regime is important. We see that \framework is better than \algname{$k$-means} when the number of clusters used is larger than the true number of clusters (except on \datasetname{reddit}, where we have argued above that DPC is not suitable). When the number of clusters used is larger, the quality of \algname{$k$-means} decays quickly while the quality of \framework is more robust. When the true number of clusters used is smaller than the ground truth, the quality of the two methods is similar on \datasetname{birds}, \datasetname{ImageNet}, and \datasetname{MNIST}, while \algname{$k$-means} is better on \datasetname{arxiv} and \datasetname{reddit}. 

Moreover, as mentioned in \Cref{sec:intro}, DPC variants can produce a hierarchy of clusters, which contains more information than $k$-means. Each run of $k$-means produces only a single cluster.  
Thus, to perform the experiment in \Cref{fig:varying_num_cluster}, 
in \framework we can just redo the postprocessing step (Lines~\ref{alg:framework:noise}--\ref{alg:framework:finish} of \Cref{alg:framework}), whereas for $k$-means we must rerun the algorithm from scratch for each choice of the number of clusters. For example, it takes about $4$ hours to generate \Cref{fig:varying_num_cluster} for \datasetname{arxiv} and and about $90$ hours for \datasetname{ImageNet}, whereas all datasets with \framework take less than a few minutes.

\begin{table*}[!t]
\footnotesize
\centering
 \begin{tabular}{l c c c c} 
 \toprule
 Dataset & Algorithm & Details & Time & ARI\\
 \midrule
\datasetname{S2} & \algname{fastdp} & N/A & 0.172 & 0.933\\
\datasetname{S2}& \algname{$k$-means} & $nredo=1$ & 0.005 & 0.860\\
\datasetname{S2}& \algname{$k$-means} & $nredo=50$ & 0.237 & 0.940\\
\datasetname{S2}& \framework & product center finder & 0.486 & 0.925\\
\datasetname{S2}& \framework & threshold center finder & 0.473 & 0.925\\
\datasetname{S2}& \algname{BruteForce} & threshold center finder & 0.499 & 0.925\\
\datasetname{S2}& \algname{DBSCAN} & $\epsilon= 52000, min\_pts=128$ & 0.006  &  0.877\\
 \midrule
\datasetname{Unbalanced}& \algname{fastdp} & N/A & 0.246 & 1.000\\
\datasetname{Unbalanced}& \algname{$k$-means} & $nredo=1$ & 0.003 & 0.691\\
\datasetname{Unbalanced}& \algname{$k$-means} & $nredo=50$ & 0.098 & 0.832\\
\datasetname{Unbalanced}& \framework & product center finder & 0.737 & 0.843\\
\datasetname{Unbalanced}& \framework & threshold center finder & 0.651 & 1.000\\
\datasetname{Unbalanced}& \algname{BruteForce} & threshold center finder & 0.623 & 1.000\\
\datasetname{Unbalanced}& \algname{DBSCAN} & $\epsilon=16000, min\_pts=3$ & 0.005 & 0.999989\\
 \bottomrule
\end{tabular}
\caption{Runtime and ARI score with respect to ground truth for all methods using a single thread on the small low-dimensional synthetic datasets \datasetname{S2} and \datasetname{Unbalanced}. When using the threshold center finder, we set $\delta_{\min}=102873$ for \datasetname{S2} and $\delta_{\min}=30000$ for \datasetname{Unbalanced}. We 
set $\rho_{\min}=0$ for the noise function. For $k$-means, we used $niter=20$. 
}
\label{tab:small_experiment}  
\end{table*}

\begin{figure}[ht]
 \centering
 \includegraphics[width=0.7\columnwidth]{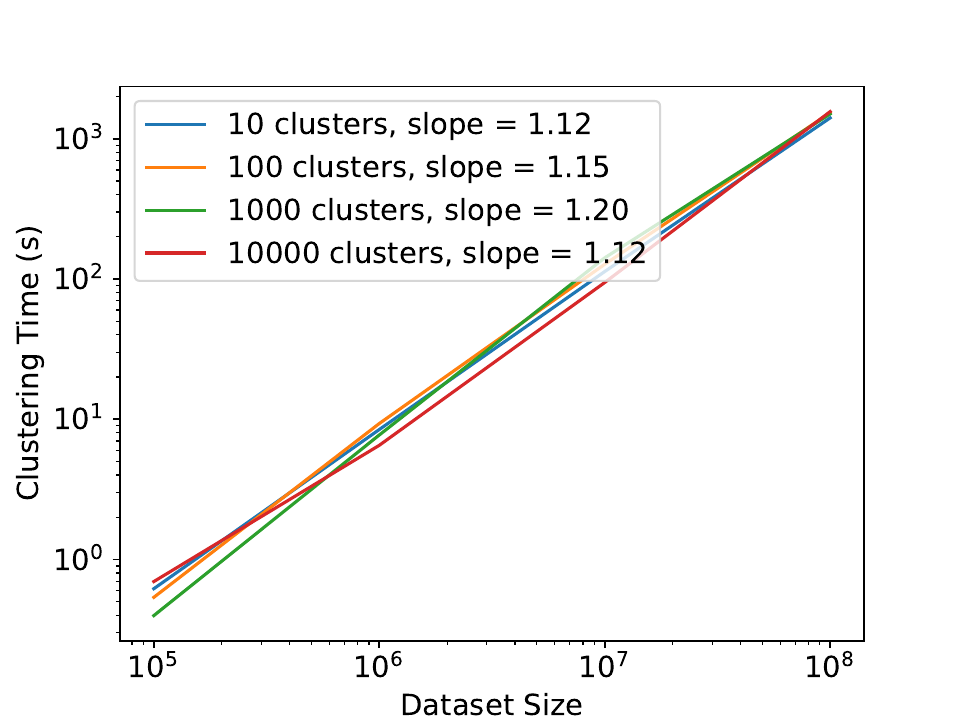}
  \caption{Running time (seconds) across \datasetname{gaussian} datasets of different sizes and different numbers of clusters in log-log scale. The "slope" $s$ is the slope of the line in the log-log plot, and means that the running time scales as $O(n^s)$. }
  \label{fig:dpc-scalability-n}
\end{figure}

\section{Details on DBSCAN Parameters}\label{sec:parameters}
In this subsection, we present the parameters we used for the DBSCAN algorithm.
Let $range(start, stop, step)$ represent the set of numbers from $start$ to $stop$ with increment $step$.
We put all noise points into a single cluster when evaluating the ARI.

We follow the guidelines for choosing parameters as described in \Cref{sec:compare-methods}. We also explored other parameters by trial and error to try our best to obtain high ARI scores using DBSCAN. However, we find that on high-dimensional data, it is difficult for DBSCAN achieve high ARI. This is consistent with the observation of the original authors of DBSCAN~\cite{schubert2017dbscan}.

For \datasetname{Unbalanced}, we used $\epsilon \in range(5000, 20000, 1000)$ and $min\_pts \in range(1, 50, 2)$. We find that the highest ARI is achieved when when $\epsilon=16000, min\_pts=3$, which gives an almost perfect clustering. There are 9 clusters with 1 noise point.

For \datasetname{S2}, we used $\epsilon \in range(40000, 70000, 2000)$ and $min\_pts \in range(100, 150, 2) \cup range(1, 50, 2)$. We find that the highest ARI is achieved when $\epsilon= 52000, min\_pts=128$. There are 16 clusters with 275 noise points.

For \datasetname{MNIST}, we used $\epsilon \in range(0.5, 9, 0.5)$ and $min\_pts \in range(1, 5, 1) \cup range(100, 1000, 200) \cup range(1500, 1700, 100) \cup range(5000, 9000, 1000)$. 
We find that the highest ARI is achieved when $\epsilon= 3, min\_pts=1$. There are 60074 clusters and no noise points. 

For \datasetname{birds}, we used $\epsilon \in range(20, 40, 5) \cup range(6, 14, 1)$ and $min\_pts \in range(2000, 2200, 100) \cup range(1, 5, 1) \cup range(120, 270, 30)$.
We find that the highest ARI is achieved when $\epsilon=12, min\_pts=1$. There are 44663 clusters and no noise points. 

On \datasetname{arxiv}, \algname{DBSCAN} with $\epsilon \geq 0.64$ runs out of memory. We used $\epsilon \in range(0.32, 0.62, 0.02)$ and $min\_pts \in range(2000, 2200, 100) \cup range(1, 5, 1) \cup [10, 50, 100, 500, 1000, 5000]$. 
We find that the highest ARI is achieved when $\epsilon=0.4, min\_pts=1$. There are 447198 clusters and no noise points. 

On \datasetname{reddit}, we used $\epsilon \in range(0.4, 0.72, 0.02)$ and $min\_pts \in range(2000, 2200, 100)  \cup  range(1, 5, 1) \cup range(3000, 13000, 1000)$.
We find that the highest ARI is achieved when $\epsilon=0.46, min\_pts=4$. There are 46132 clusters and 8089 noise points.

On \datasetname{ImageNet}, \algname{DBSCAN} with $\epsilon \geq 36$ runs out of memory. We used $\epsilon \in range(20,34,1)$ and $min\_pts \in range(2000, 2200, 100) \cup range(1, 5, 1) \cup range(700, 1300, 200)$. 
We find that the highest ARI is achieved when $\epsilon=20, min\_pts=700$. There are 393 clusters and 606651 noise points.